\documentclass[10pt,preprint,a4paper]{aastex}

\usepackage{amsmath}                
\usepackage{amsfonts}               
\usepackage{amssymb}                
\usepackage{epsfig}                 

\usepackage{subfigure}

\usepackage{graphicx}
\usepackage{grffile}

\newcommand{\kms}{{~\rm km\; s^{-1}}}

\newcommand{\cm}{{~\rm cm}}
\newcommand{\s}{{~\rm s}}
\newcommand{\km}{{~\rm km}}
\newcommand{\g}{{~\rm g}}
\newcommand{\K}{{~\rm K}}
\newcommand{\erg}{{~\rm erg}}
\newcommand{\yr}{{~\rm yr}}
\newcommand{\Myr}{{~\rm Myr}}

\newcommand{\kpc}{{~\rm kpc}}

\begin{document}

\title{HEATING THE INTRA-CLUSTER MEDIUM BY JET-INFLATED BUBBLES}

\author{Shlomi Hillel\altaffilmark{1} and Noam Soker\altaffilmark{1}}

\altaffiltext{1}{Department of Physics, Technion -- Israel Institute of Technology, Haifa
32000, Israel; shlomihi@tx.technion.ac.il, soker@physics.technion.ac.il}

\begin{abstract}
We examine the heating of the intra-cluster medium (ICM) of
cooling flow clusters of galaxies by jet-inflated bubbles and
conclude that mixing of hot bubble gas with the ICM is
more important than turbulent heating and shock heating.
We use the {\sc pluto} hydrodynamical
code in full 3D to properly account for the inflation of the
bubbles and to the multiple vortices induced by the jets and
bubbles. The vortices mix some hot shocked jet gas with the ICM.
For the parameters used by us the mixing process accounts for about four times as much heating as that by the kinetic energy in the ICM, namely, turbulence and sound waves.
We conclude that turbulent heating plays a
smaller role than mixing. Heating by shocks is even less efficient.
\end{abstract}


\section{INTRODUCTION}
\label{s-introduction}

A negative feedback mechanism determines the thermal evolution of
the intra-cluster medium (ICM) in the inner regions of cooling
flow (CF) clusters and groups of galaxies (\citealt{McNamaraNulsen2007, McNamaraNulsen2012}; some examples include \citealt{Farage2012, Pfrommer2013}).
This feedback mechanism is
driven by active galactic nucleus (AGN) jets that inflate X-ray
deficient cavities (bubbles; e.g., \citealt{Dong2010,
OSullivan2011, Gaspari2012a, Gaspari2012b, GilkisSoker2012}).
Examples of bubbles in cooling flows include Abell 2052
\citep{Blanton2011}, NGC 6338 \citep{Pandge2012}, NGC 5044
\citep{David2009}, and HCG 62 \citep{Gitti2010}, among many
others.

Vortices inside the bubbles and in their surroundings play major
roles in the formation of bubbles, their evolution, their
interaction with the ICM, and the dynamics and formation of cold
regions (e.g. \citealt{Omma2004, Heinz2005, Roediger2007,
Sternberg2008b, GilkisSoker2012}). Some cool regions further cool
and flow inward to feed the AGN. The process of feeding the AGN
with cold clumps in the feedback mechanism cycle is termed the
\emph{cold feedback mechanism}, and was suggested by
\cite{Pizzolato2005}. The cold feedback mechanism has been later
strengthened by observations of cold gas and by more detailed
studies (e.g., \citealt{Revaz2008, Pope2009, Pizzolato2010,
Edge2010, Wilman2011, Nesvadba2011, Cavagnolo2011, Gaspari2012a,
Gaspari2012b, McCourt2012, Sharma2012, Farage2012, Waghetal2014,
BanerjeeSharma2014, McNamaraetal2014, VoitDonahue2015,
Voitetal2015, Lietal2015, Prasadetal2015, Russelletal2015, Tremblayetal2015, Fogartyetal2015}).

\cite{GilkisSoker2012} and \cite{HillelSoker2014} have shown that
the mixing of the hot bubble gas with the ICM is a much more
efficient heating process than heating by repeated shocks; hereafter termed shocks-heating.
As the axis of the
bubbles changes with time and/or the central jets' source and the
ICM move relative to each other, e.g., the galaxy group NGC 5813
\citep{Randalletal2015}, over time the mixing-heating operates in
all directions. Shocks-heating is extremely inefficient, and is
unlikely to explain the feedback mechanism \citep{Sokeretal2013}.
\cite{Randalletal2015} argue for shocks-heating in NGC~5813,
ignoring the counter arguments for shocks-heating in NGC~5813
\citep{Sokeretal2013}. We hold that even in NGC~5813 mixing is
more important than shocks in heating the ICM
\citep{Sokeretal2015}. This heating by mixing closes the cycle in
the cold feedback mechanism.

A key to the study of the mixing heating is to inflate bubbles
self-consistently. This requires either slow (sub-relativistic)
massive wide (SMW) jets
\citep{Sternberg2007}, precessing jets \citep{Sternberg2008a,
Falceta-Goncalves2010}, or a relative motion of the jets to the
medium \citep{Bruggen2007, Soker2009, Morsony2010, Mendygral2012}.
In the present study we inflate bubbles by SMW jets, also
supported by observations (e.g., \citealt{Moe2009, Dunn2010,
Aravetal2013}), but our results hold for bubbles inflated by
precessing jets or a relative motion of the ICM as well.

In a recent paper \cite{Zhuravlevaetal2014} claim that heating by
turbulence dominates in the Perseus cooling flow cluster. We here
compare the energy that jet-inflated bubbles channel to turbulence
and to the direct thermal energy of the ICM.
The axi-symmetric 2D simulations of \cite{GilkisSoker2012} and \cite{HillelSoker2014}
are insufficient to account for the vortices and turbulent motion,
which are expected to be fully three-dimensional.
Therefore, as 2D axi-symmetry restricts the motion and changes the character of the turbulence invoked,
for the goal of this paper, that is to study the role of turbulent heating, we lift this restriction and employ 3D simulations.

The 3D numerical setting is described in section \ref{s-numerical-setup}. In
sections \ref{s-flow-structure} we present the flow structure, and
in section \ref{s-heating-the-icm} we study the heating of the
ICM. Our short summary is in section \ref{s-summary}.

\section{NUMERICAL SETUP}
\label{s-numerical-setup}

We use the {\sc pluto} code \citep{Mignone2007} for the
hydrodynamic simulations, in a three-dimensional Cartesian grid
with adaptive mesh refinement (AMR). The computational grid is in
the octant where the three coordinates $x$, $y$ and $z$ are
positive. At the $x = 0$, $y = 0$ and $z = 0$ planes we apply
reflective boundary conditions. The $z$ coordinate is chosen along
the initial axis of the jets. In reality two opposite jets are
launched simultaneously, such that the flow here is assumed to be
symmetric with respect to the $z = 0$ plane, amounting to
reflective boundary conditions at $z = 0$. The base computational
grid (lowest AMR level) spans the cube $0 \leq x, y, z \leq 50 \kpc$, with $16$
divisions in each direction.
Up to $5$ AMR levels are employed with a refinement
ratio of $2$.
Thus, the highest resolution is $\approx 0.1 \kpc$.
The refinement criterion is the default AMR criterion in {\sc pluto} v. 4.1,
based on the second derivative error norm \citep{Lohner1987} of the total energy density.

Heat conduction and viscosity are not included in the simulations.
However, local heat conduction is expected to be efficient on scales of $\lesssim 0.1 \kpc$ (\citealt{Soker2010}),
and therefore is effectively incorporated via the resolution of the simulations
and does not need to be included. The conclusions are found not to be sensitive to the resolution.
Magnetic fields, also not included in the current numerical study, would in reality prevent global heat conduction, but not local heat conduction.
The vortices would entangle field lines, leading to local magnetic field line reconnection and hence allowing local heat conduction.

In this study we examine the increase in the kinetic energy of the ICM.
The kinetic energy results from turbulence, sound waves, and gas motion induced by cavity motion, i.e., cavity heating \citep{Nulsenetal2007}.
The dissipation of turbulence (e.g., \citealt{Zhuravlevaetal2014}) and sound waves (e.g., \citealt{ShabalaAlexander2007}) heats the gas.
We do not include dissipation, but we calculate the increase in the kinetic energy of the ICM and term it in short turbulence.
However, it should be kept in mind that it actually includes the energy of sound waves and uplifted gas, hence we do account for sound wave heating and cavity heating.
If the dissipation is not $100\%$ efficient in transferring kinetic to thermal energy, then we are overestimating this turbulent energy source for heating. 
The heating by shock waves is examined directly, as shock dissipation is treated by the numerical code.  
Heating by cosmic rays, e.g., \cite{GuoOh2008}, are not included in our simulations at all. 

On the outer boundaries we use an outflow boundary
condition.
At the boundary $z = 0$ we inject into the grid a jet with a
half-opening angle of $\theta_{\rm j} = 70^\circ$
\citep{Sternberg2007}. The jet material is inserted through a
circle $x^2 + y^2 \leq r_{\rm j}^2$ on the $z = 0$ plane with a
radius of $r_{\rm j} = 3 \kpc$. The initial jet velocity in the nominal case is $v_{\rm j} =
8200 \kms$, a Mach number of about $10$. The direction of
the velocity at each injection point $(x,y,0)$ in the circle is
$\hat{v} = (x, y, h_{\rm j}) / \sqrt{x^2 + y^2 + h_{\rm j}^2}$,
where $h_{\rm j} = r_{\rm j} / \tan \theta_{\rm j}$. The jet is
injected during each active episode, and when the jet is turned
off reflective boundary conditions apply in the whole $z = 0$
plane.
In the nominal run, the jet is periodically turned on for $10\Myr$ and off for $10\Myr$.
The power of the two jets together is
\begin{equation}
\label{eq: jet power}
\dot E_{2{\rm j}} = 2 \times 10^{45} \erg \s^{-1},
\end{equation}
half of it in each direction. The
mass deposition rate is thus
\begin{equation}
\dot{M}_{2{\rm j}} = \frac{2 \dot E_{2{\rm j}}}{v_{\rm j}^2} = 94
M_{\odot}~\yr^{-1}.
\end{equation}

In two additional simulations we vary two parameters.
In one simulation, termed Run B below, we reduce the jet activity and quiescence time intervals by a factor of $5$.
In a different simulation, Run C, we increase the velocity by a factor of $\sqrt{10}$ and lower the mass deposition rate, keeping the jet power unchanged.

As we do not include magnetic fields, hence their influence on instabilities, the roles of magnetic fields in the mixing process and in small scale motions are not explored here.
Although in two dimensions magnetic fields act to stabilize some instabilities (e.g., \citealt{Robinsonetal2004}), in three dimensions the magnetic fields cannot suppress instabilities with wave-vector components perpendicular to the field lines.
Hence, we do not expect magnetic fields to suppress mixing, but only to influence the details of the mixing.
Another effect is that the violent vorticity we find can amplify magnetic fields to the point that they start to reconnect and heat the gas.
This overall process channels kinetic energy to magnetic energy and then to thermal energy, hence increasing the heating efficiency of the mixing process. 
Overall we do not expect magnetic fields to influence the main conclusions of this study.

The simulation begins with an isothermal box of gas at an initial
temperature of $T_{\rm ICM} (0) = 3 \times 10^7 \K$ with a density
profile of (e.g., \citealt{VernaleoReynolds2006})
\begin{equation}
\rho_{\rm ICM}(r) = \frac{\rho_0}{\left[ 1 + \left( r / a \right)
^ 2 \right] ^ {3 / 4}},
\end{equation}
with $a = 100 \kpc$ and $\rho_0 = 10^{-25} \g \cm^{-3}$. A gravity
field is added to maintain an initial hydrostatic equilibrium, and
is kept constant in time.
We include radiative cooling in the simulations, where the tabulated cooling function is taken from the solar-metalicity values of Table 6 in \cite{SutherlandDopita1993}.

We run simulations with either $4$ or $5$ levels of AMR, to check
for numerical convergence. We found no noticeable differences in
the results.  We also run simulations with a twice as large grid
and no reflecting boundary conditions on the plane $x=0$. Namely,
the $x$-coordinate covers the range $-50 \kpc \leq x \leq 50
\kpc$. We found no significant changes in the results, hence
conclude that the reflecting boundary conditions do not affect our
results, both qualitatively and quantitatively.

\section{FLOW STRUCTURE}
\label{s-flow-structure}

\begin{figure}[!htb]
\centering
\subfigure{\includegraphics[width=0.45\textwidth]{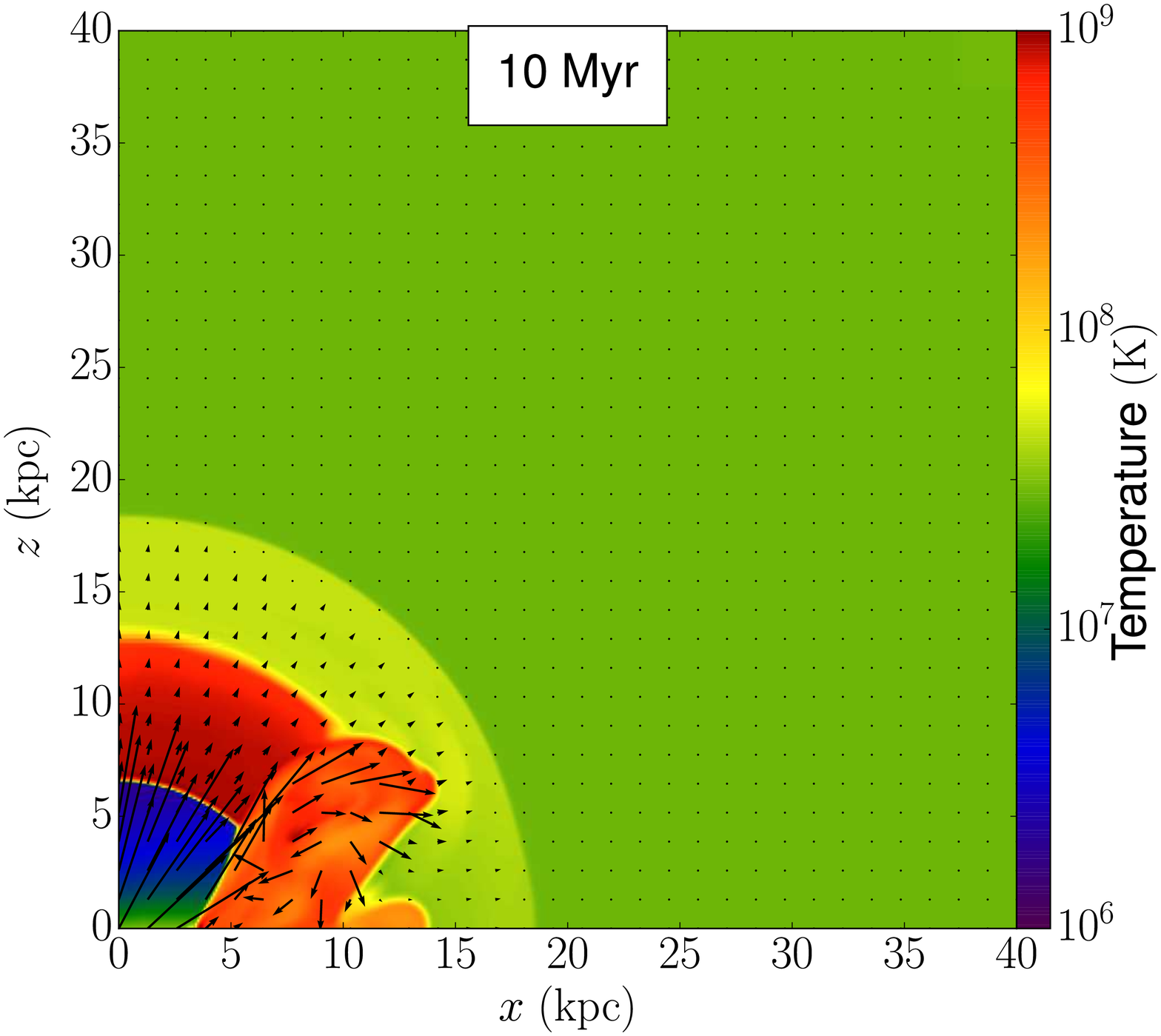}}
\subfigure{\includegraphics[width=0.45\textwidth]{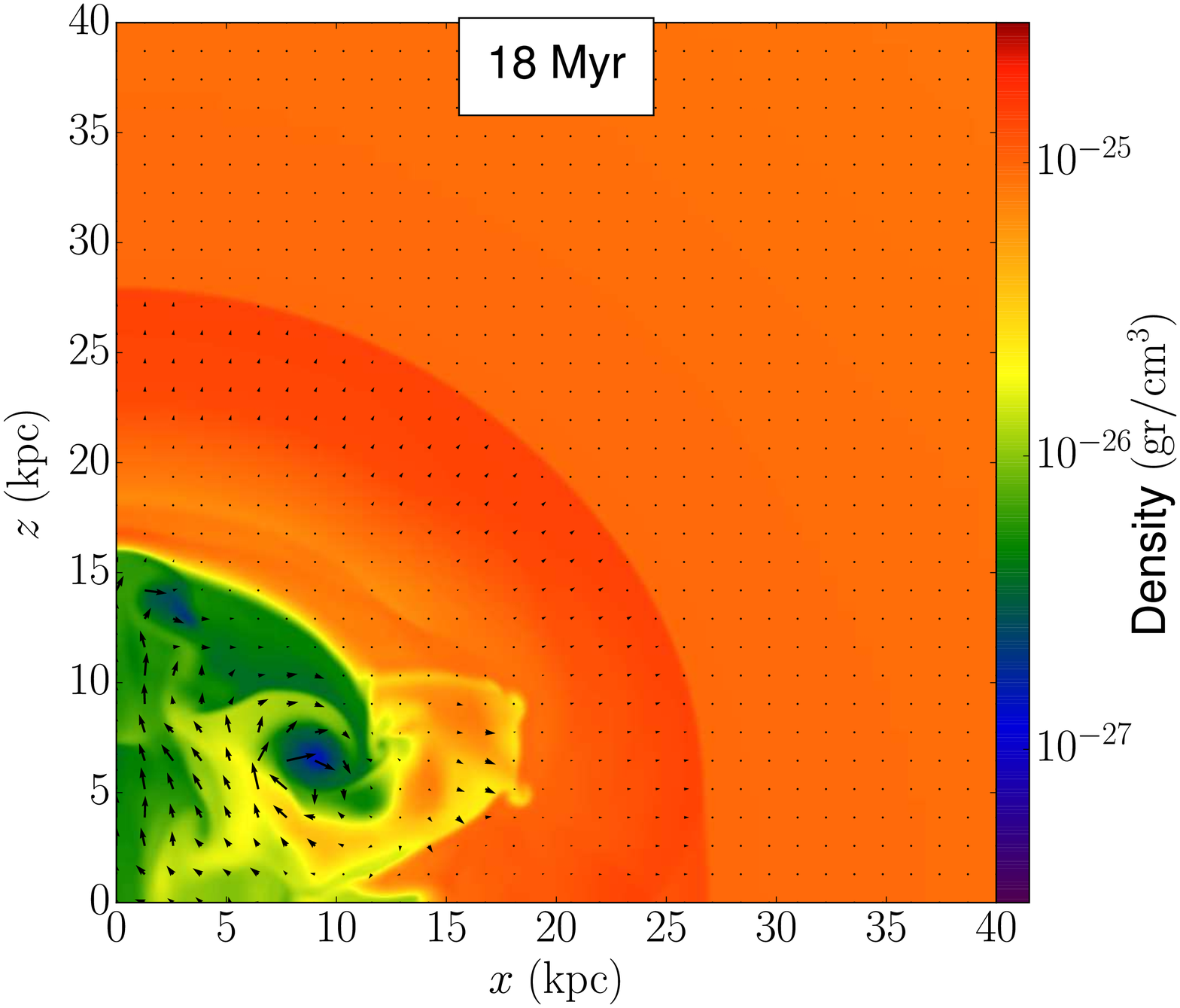}}\\
\subfigure{\includegraphics[width=0.45\textwidth]{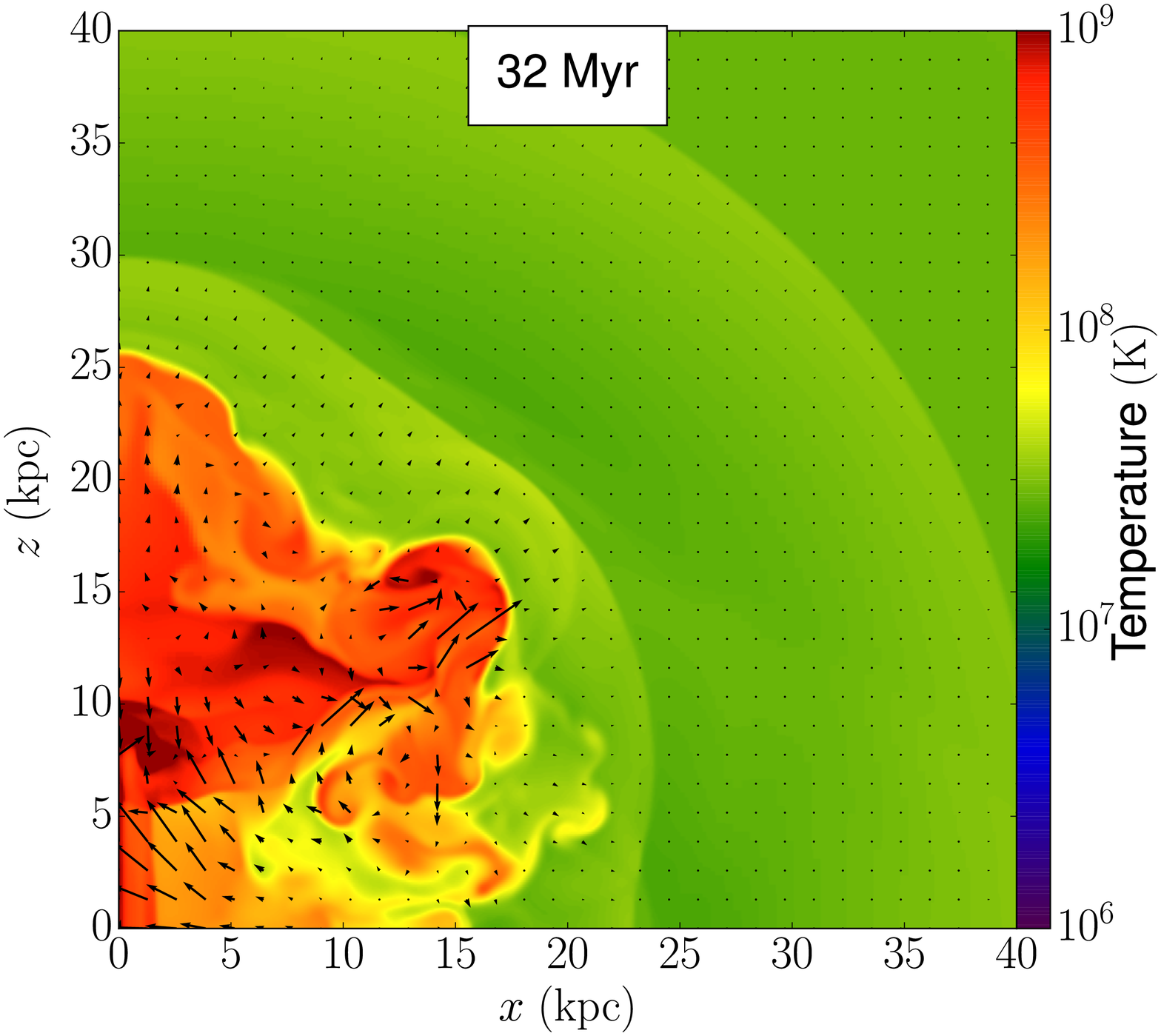}}
\subfigure{\includegraphics[width=0.45\textwidth]{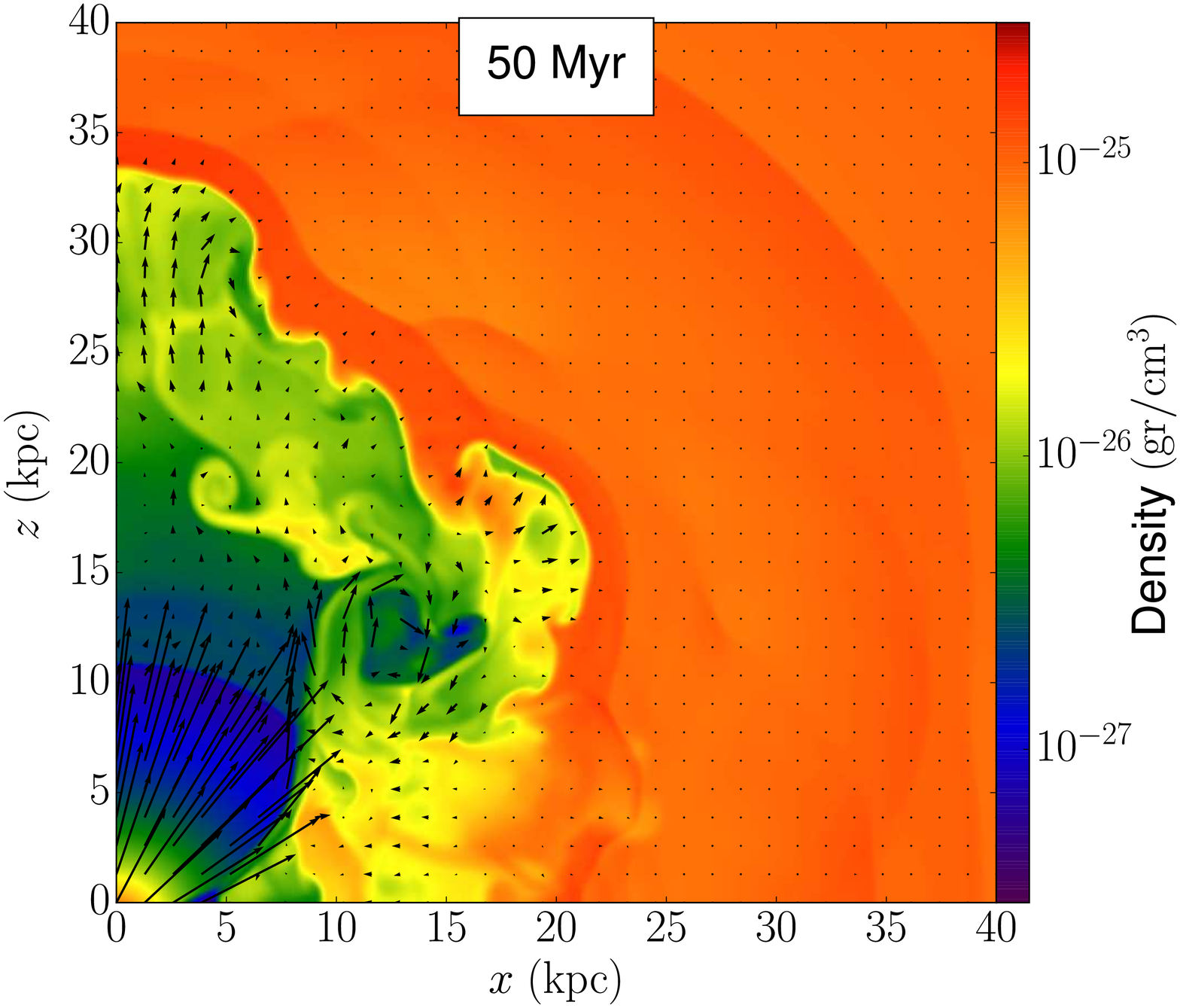}}
\caption{Evolution of the flow for the nominal case parameters given in section
\ref{s-numerical-setup}, presented in the meridional plane $y = 0$
at four times. The jet is injected through a circle of radius
$r_{\rm j} = 3 \kpc$ on the $z = 0$ plane. The half-opening angle
is $\theta_{\rm j} = 70^\circ$, i.e., the jet velocity on the
boundary of the circle has an angle of $70^\circ$ with respect to
the $z$-axis. The left panels present temperature maps and the
right panels mass density maps.  The color coding of temperature and density is
in logarithmic scale. Arrows show the velocity, with length
proportional to the velocity magnitude. A length of $1 \kpc$ on
the map corresponds to $1700 \km \s^{-1}$. When the jet is active,
the length of arrows close to the origin corresponds to $8200 \kms$. }
\label{figure: flow structure}
\end{figure}

We begin by presenting in Fig.~\ref{figure: flow structure} the
flow structure at different times.
This high-resolution simulation was run for $50\Myr$. To
follow the evolution of the same setting for $240 \Myr$ we used a
lower resolution grid. In the common time the results of the two
runs are similar.

As evident from Fig.~\ref{figure: flow structure}, e.g., at $t=32
\Myr$, we manage to inflate bubbles which match those seen in
observations as X-ray deficient cavities \citep{Sternberg2007,
HillelSoker2014, GilkisSoker2012}. The inflation of bubbles
similar to those observed in a self-consistent manner is crucial
to the study of the different heating mechanisms: mixing, shocks,
and the excitation of ICM turbulence that can dissipate later on.

\begin{figure}[!htb]
\centering
\subfigure{\includegraphics[width=0.45\textwidth]{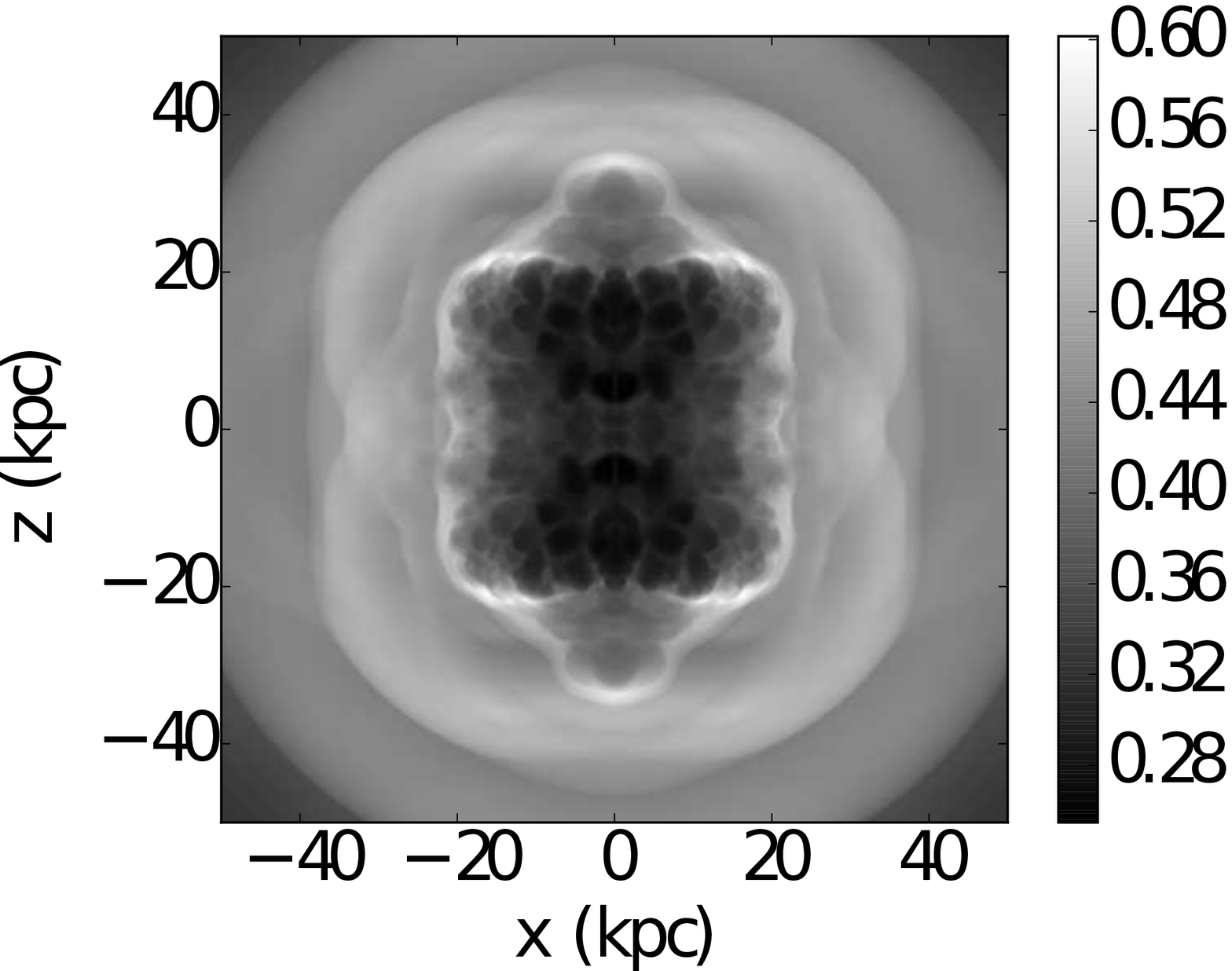}}
\subfigure{\includegraphics[width=0.45\textwidth]{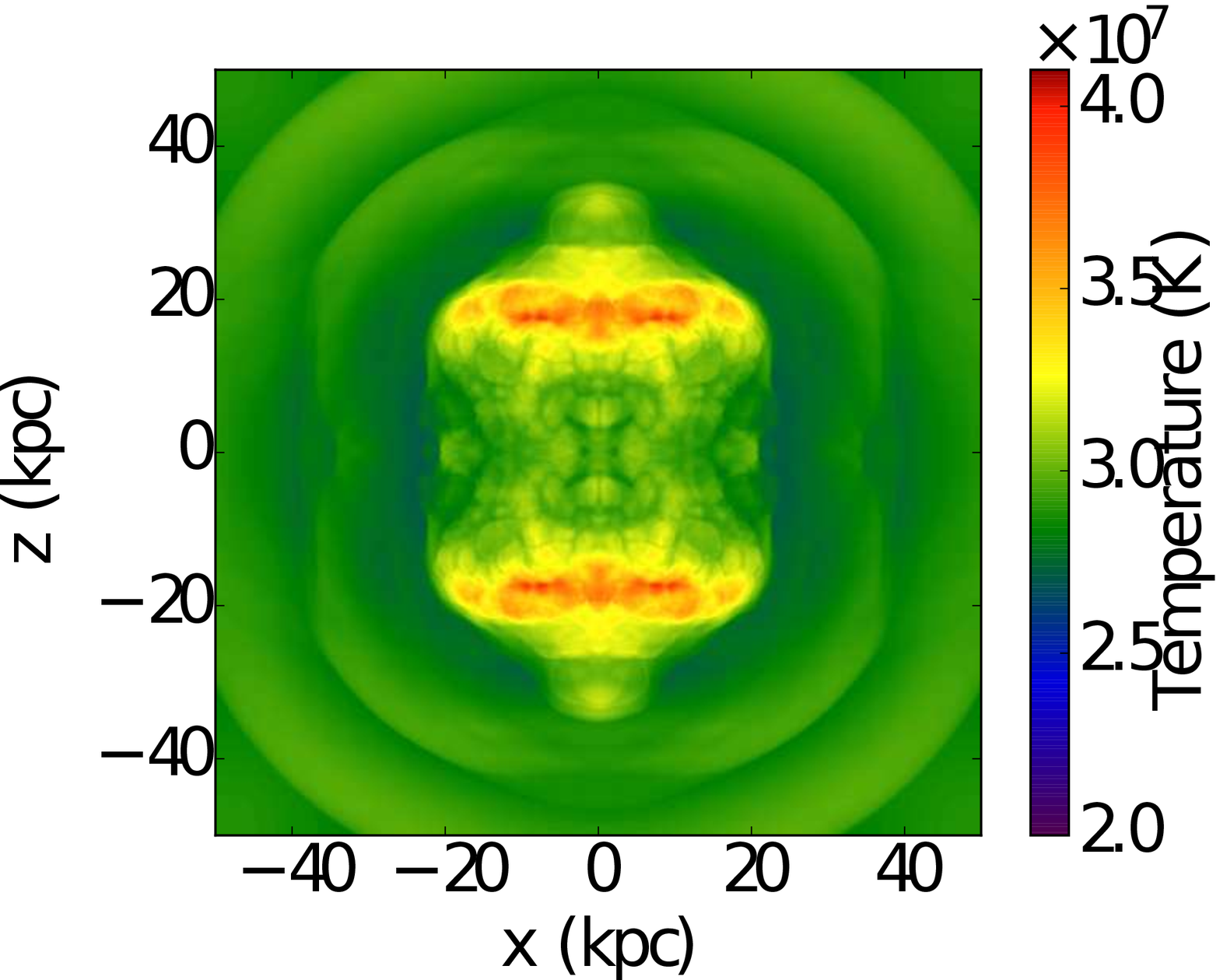}}
\caption{
Mock X-ray image (left panel) and emission-weighted temperature map (right panel) of the nominal simulation at time $t=50\Myr$.
The simulated octant was folded twice to create these panels.
Hence the four quarters about the origin are identical to each other. 
The X-ray image is created by the integration of the density squared along lines parallel to the $y$-axis.
The units are arbitrary.
The emission-weighted temperature map is made by calculating the average, along the line of sight,
of the temperature weighted by the emission power, Eq.~(\ref{eq: emission-weighted temperature}).
Note the X-ray deficient bubbles inflated by the jet.
}
\label{figure: imaging}
\end{figure}

In order to compare to observational data, in the left panel of Fig.~\ref{figure: imaging} we present mock X-ray images of the nominal simulation at time $t=50\Myr$.
This image was created by folding the simulated octant twice, such that the four quarters of the plane are identical.
Although the initial setup of the simulation is cylindrically symmetric,
hydrodynamic instabilities magnified numerically cause departures from that symmetry.
In particular, the surface of the low density hot bubble is ripply and uneven, with several relatively dense filaments crossing that volume near the bubble's surface
(later discussed in relation to Fig.~\ref{figure: additional simulations}).
These appear as a net of bright filaments covering the dark (low density) bubble on the left panel of Fig.~\ref{figure: imaging}.  
Present X-ray observations are not sharp enough to detect such filaments.
In addition, more sophisticated simulations might have erased these filaments by including realistic local heat conduction. 

The `waist' between the two bubbles in the plane $z=0$ obtained here is wider than in typical observed bubbles. 
Our method of injecting the jet, through a circle on the surface $z=0$, is one of the causes of the wide waist near $z=0$.
In addition, on the plane $z=0$ we have reflective boundary conditions which add to the numerical error near the plane.
However, overall, we obtain an image which reasonably resembles observations.

The right panel of Fig.~\ref{figure: imaging} is an emission-weight temperature map at the same time, $t=50\Myr$.
The emission-weighted temperature at each point $(x,z)$ in the image is the average of the temperature, along the line of sight ($y$-axis here),
weighted by the emission power,
\begin{equation}
\label{eq: emission-weighted temperature}
T_{\rm ew} = \frac{\int{\Lambda(T) n^2 T {\rm d}y}}{\int{\Lambda(T) n^2 {\rm d}y}}.
\end{equation}

For reference we also provide representative shock Mach number values.
At the front of the jet, on the $z$-axis, the first shock wave running through the ICM has a Mach number of $\mathcal{M} = 1.53$
at $z=20\kpc$ and a Mach number of $\mathcal{M} = 1.18$ when it reaches a distance of $z=40\kpc$.
As we show later, such weak shocks are insignificant in heating the ICM.

To better capture some properties of the flow we present
artificial flow quantities called `tracers'. The tracers are
frozen-in to the flow, and hence represent the spreading with
time of gas starting in a certain volume. A tracer's initial value
is set to $\xi (0) = 1$ in a certain volume and $\xi (0) = 0$
elsewhere. If the traced gas is mixed with the ICM or the jet's
material, its value drops to $0 < \xi(t) < 1$.

\begin{figure}[!htb]
\centering
\subfigure{\includegraphics[width=0.45\textwidth]{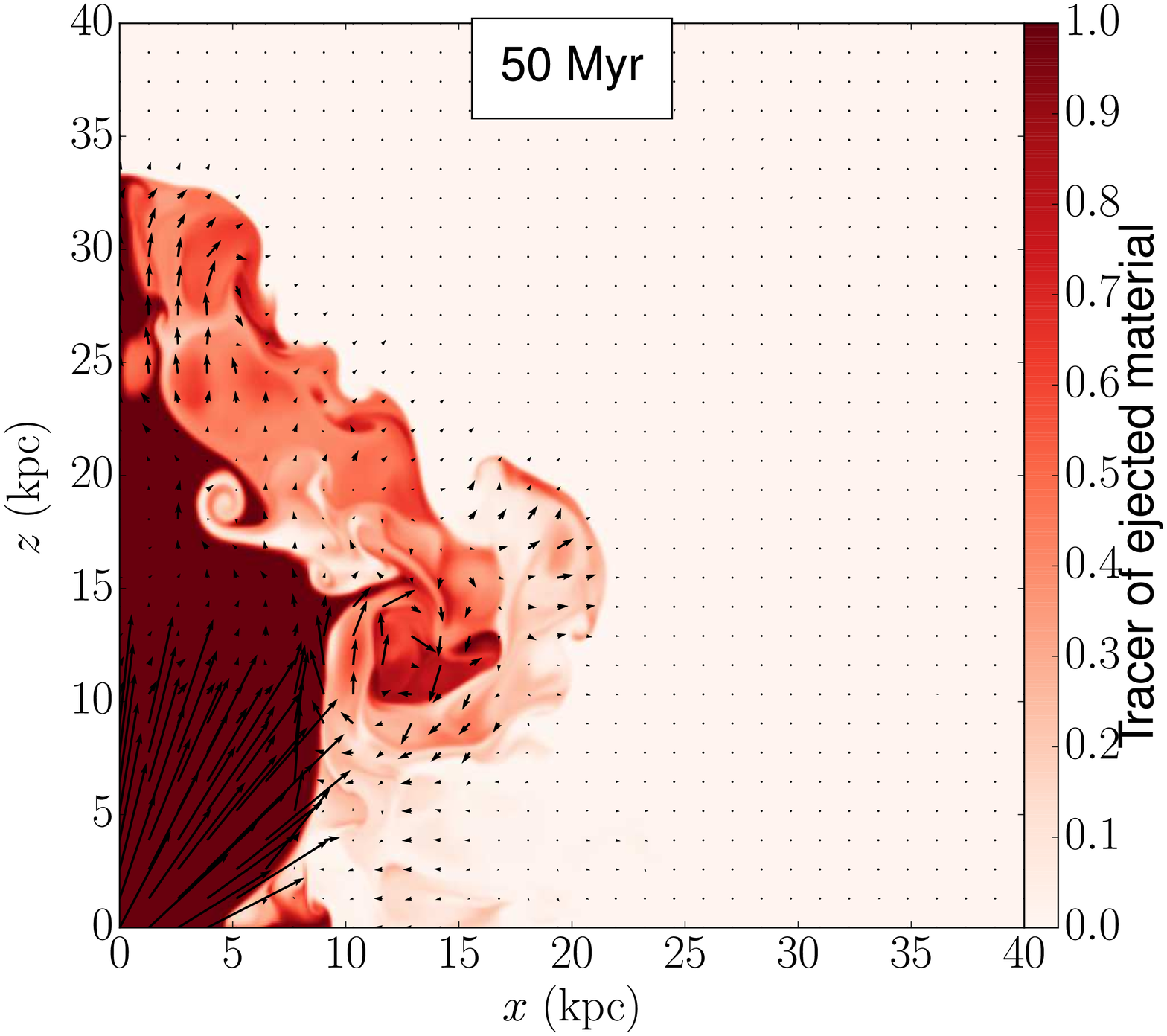}}
\subfigure{\includegraphics[width=0.45\textwidth]{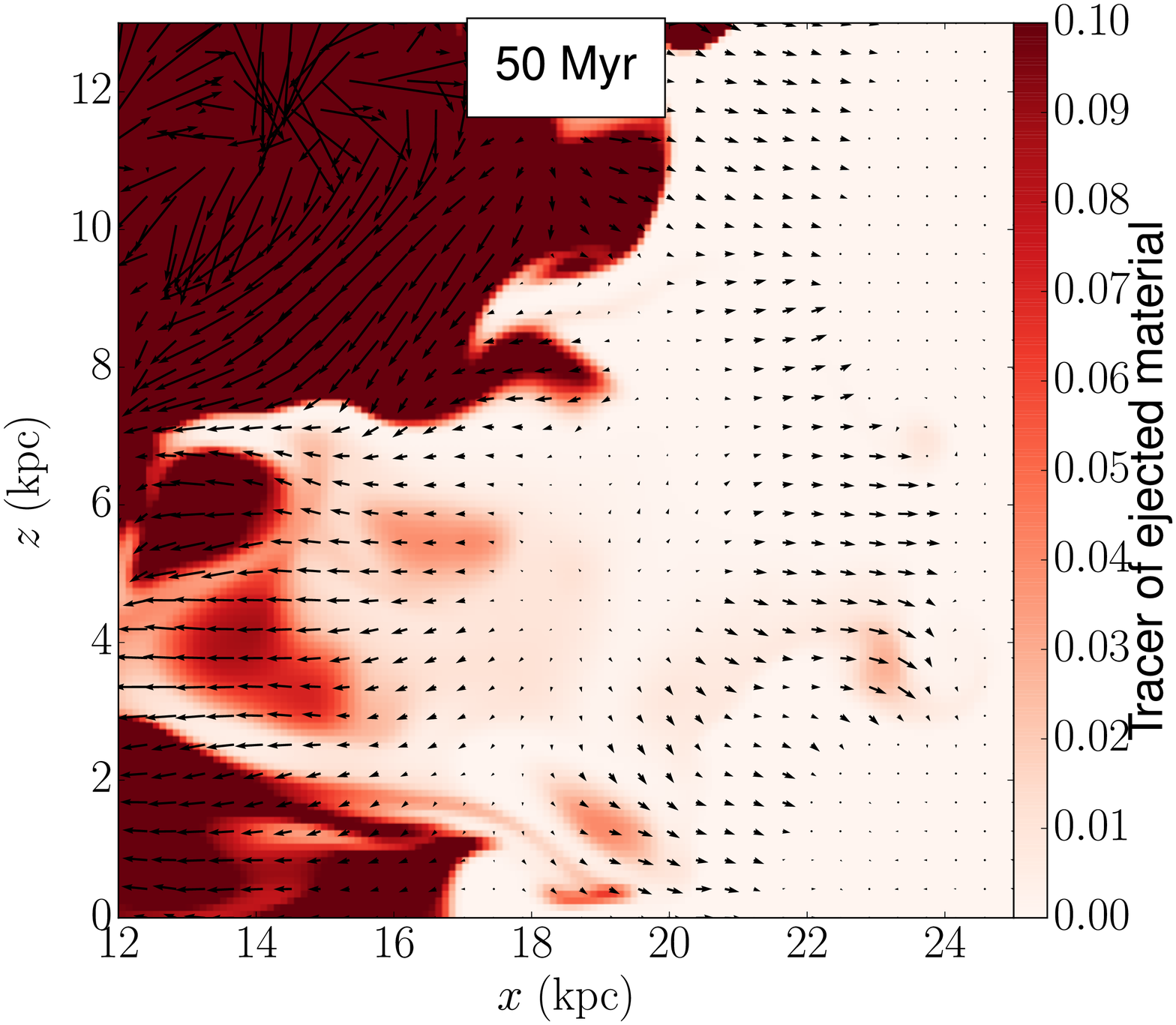}}\\
\subfigure{\includegraphics[width=0.45\textwidth]{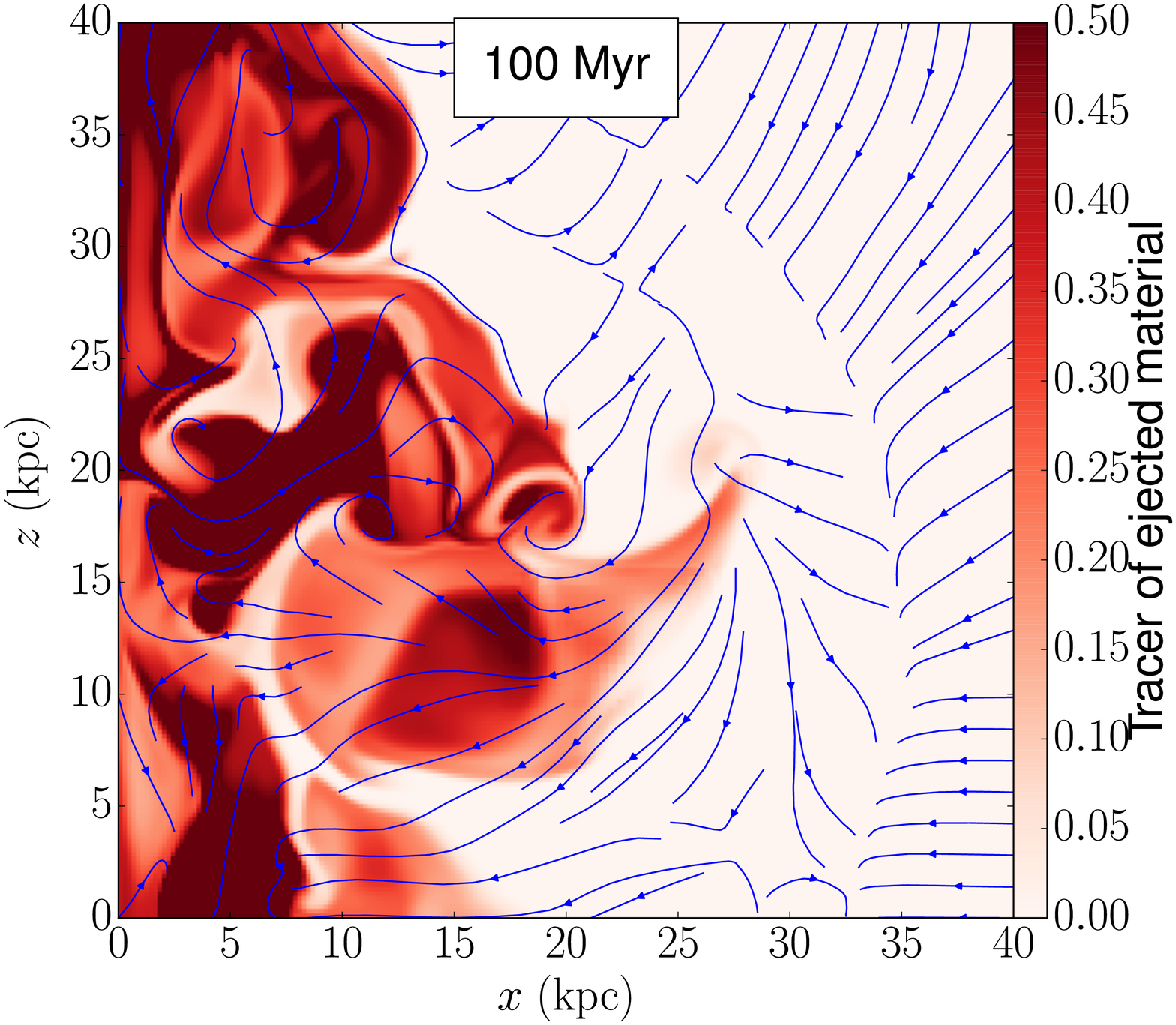}}
\subfigure{\includegraphics[width=0.45\textwidth]{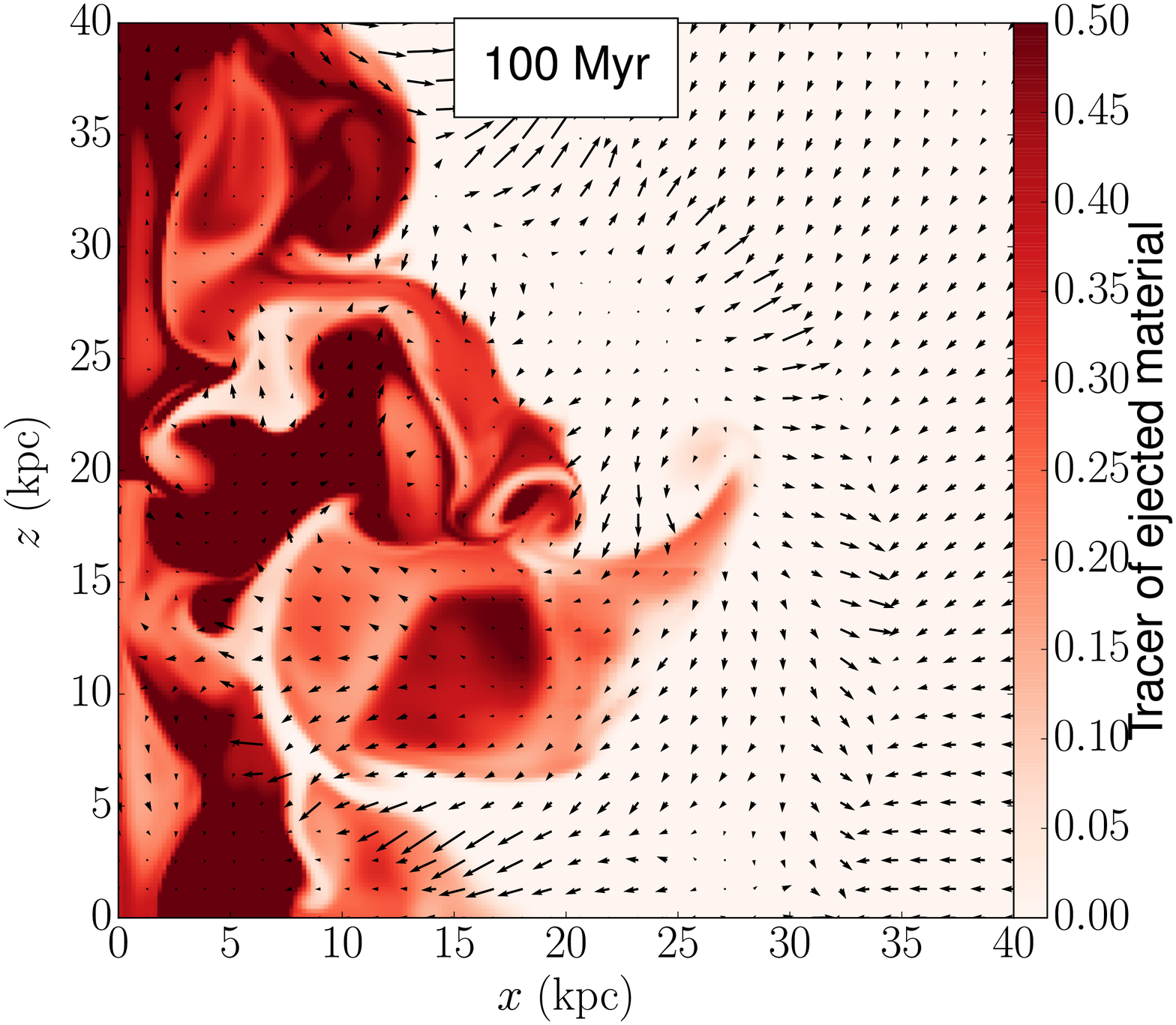}}
\caption{ A $y = 0$ slice of the tracer that is frozen-in to the
jet material, at times $t = 50$ and $100 \Myr$ in a simulation
with the same parameters as in Fig.~\ref{figure: flow structure}.
The value of the tracer is color-coded according to the scale
shown on the right of each panel. In two panels the length of
the arrow is linear with velocity up to an upper limit $v_{\rm m}$. In
the upper-left panel the largest velocity arrow corresponds to
$v_{\rm m}=8200\kms$, the initial jet velocity. Faster regions are
presented with an arrow length corresponding to $v_{\rm m}=8200\kms$.
The upper-right panel zooms on a region away from the center to
emphasize mixing of jet's material away from the center and near
the equatorial plane. The largest velocity arrow corresponds to
$v_{\rm m} = 3000\kms$.  In the lower-left panel we show stream lines of
the flow at $t = 100 \Myr$, and in the lower-right panel we show
mass flux arrows, i.e., $\phi=v \rho$. A length of $1 \kpc$ on the
map corresponds to $\phi = 2.3 \times 10^{-23} \km \s^{-1} \g
\cm^{-3}$.  }
 \label{figure: tracers tr1}
\end{figure}

In Fig.~\ref{figure: tracers tr1} we present maps of a tracer that
follows the ejected jet material. In the simulation the jet is
injected via the boundary condition at $z = 0$ (see section
\ref{s-numerical-setup}), and the injected material is marked with
a tracer that follows its evolution throughout the simulation. We
plot the tracer distribution in the $y = 0$ meridional plane at two times as
indicated in the different panels. Each point in space in which
the tracer value is $\xi> 0$ indicates that material from the jet
has reached it. We find again that a prominent phenomenon in such
simulations of self-consistent bubble inflation is the appearance
of vortices and mixing \citep{GilkisSoker2012, HillelSoker2014}.
The inflation of the bubble by the jet induces a complicated flow
structure with multiple vortices, some induced by vortex shedding
\citep{RefaelovichSoker2012, Walgetal2013}.

In the upper-left left panel of Fig.~\ref{figure: tracers tr1} we
present the entire range of $0 \le \xi \le 1$, while in the
upper-right panel we zoom on a specific region and emphasize the
range $0 < \xi < 0.1$. The arrows represent the direction and
magnitude of velocity. Too long arrows are truncated with a length
corresponding to a maximum velocity $v_{\rm m}$ as explained the
caption. Vigorous mixing caused by vortices can be seen in these
two panels, i.e., the vortex at $(x,z)=(5,18) \kpc$. In the
upper-right panel we can notice jet's material that reaches the
point $(x,z)=(23,3.5) \kpc$. This clearly shows that the mixing of
hot shocked jet's gas can heat the ICM to large distances in
directions perpendicular to the initial jet direction. Considering
that in a more realistic scenario the jets precess and eject
material in other directions, the mixing is more efficient even
than what is found here.

The two lower panels of Fig.~\ref{figure: tracers tr1} present a
later evolution time. The curved arrows in the lower-left panel
represent stream lines, and the arrows in the lower-right panel
represent mass flux $\phi = v \rho$. Both the stream lines and the
arrows show a shock expanding outward at $r \approx 35 \kpc$. We
can see again jet's material that reaches regions away from the
jets' axis, here at $(x,z)=(28,20) \kpc$.

\begin{figure}[!htb]
\centering
\subfigure{\includegraphics[width=0.45\textwidth]{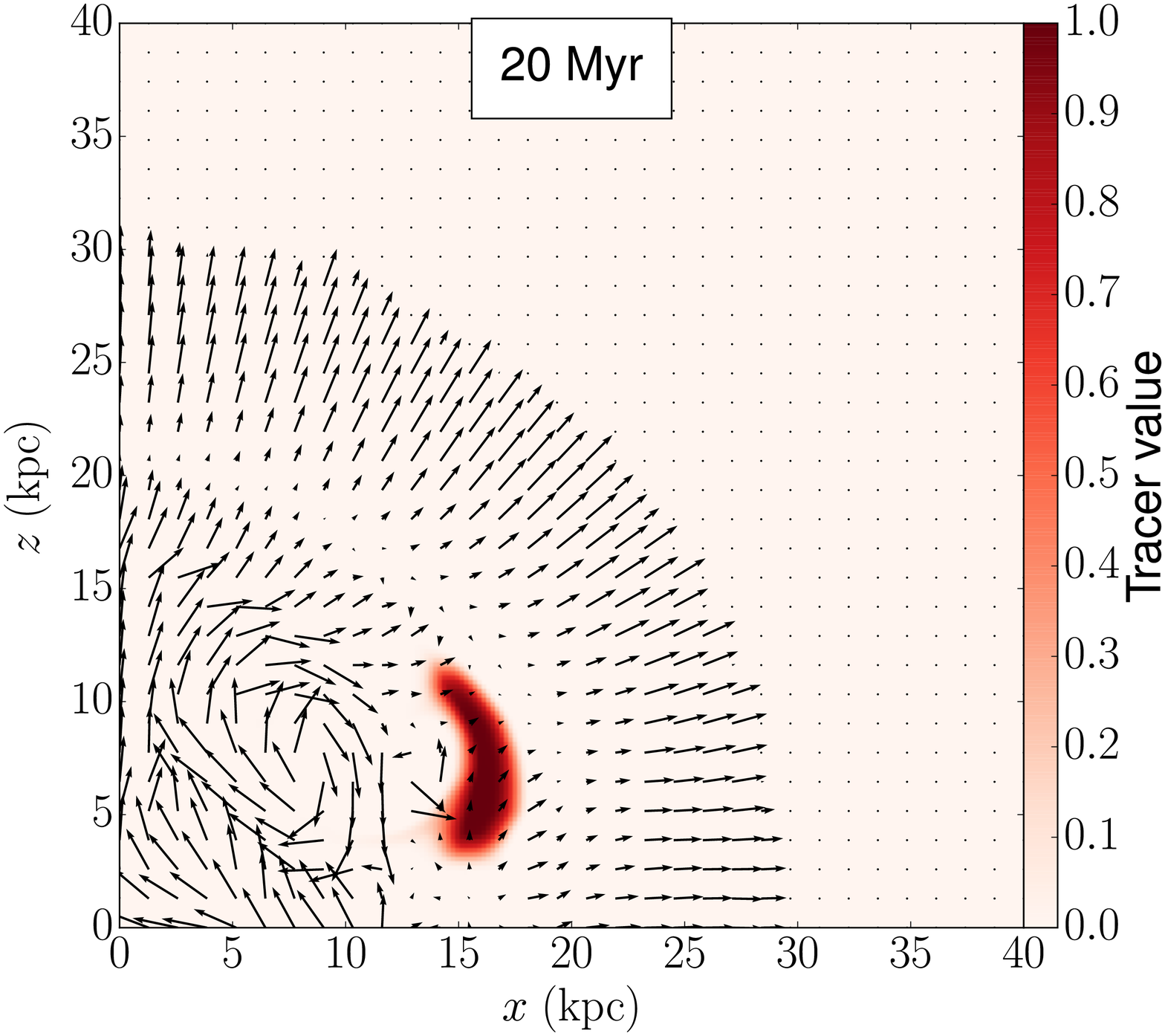}}
\subfigure{\includegraphics[width=0.45\textwidth]{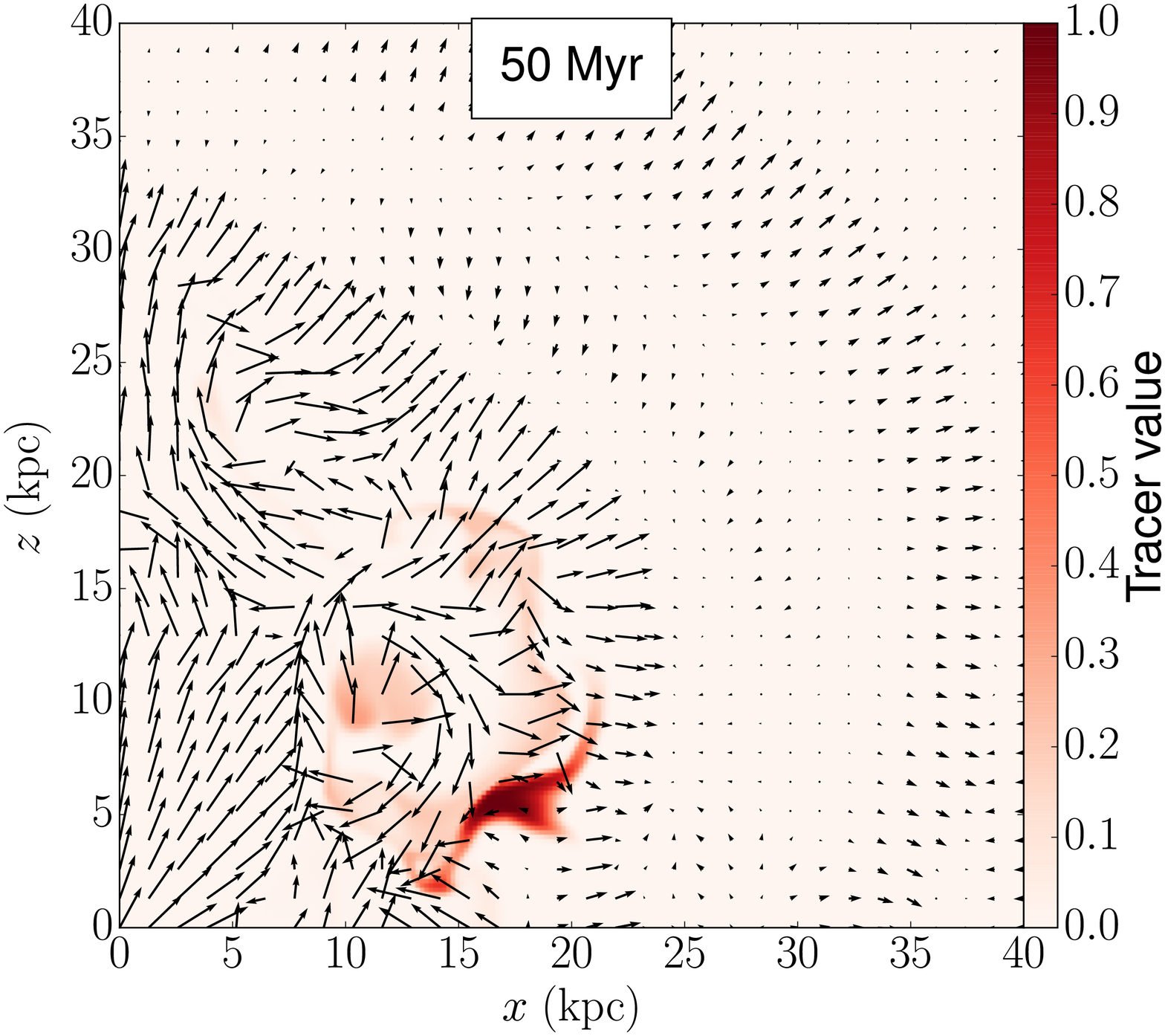}}\\
\subfigure{\includegraphics[width=0.45\textwidth]{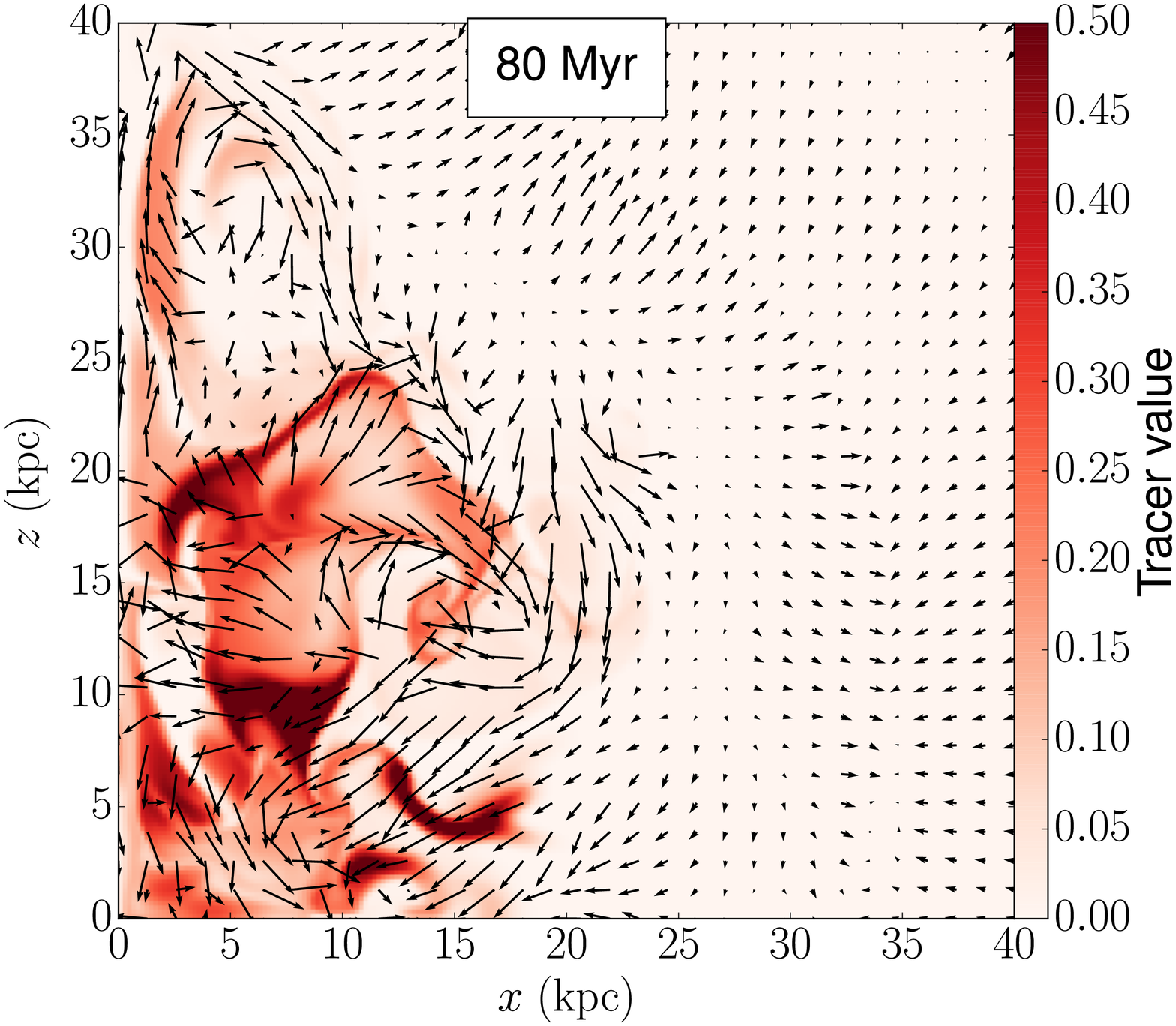}}
\subfigure{\includegraphics[width=0.45\textwidth]{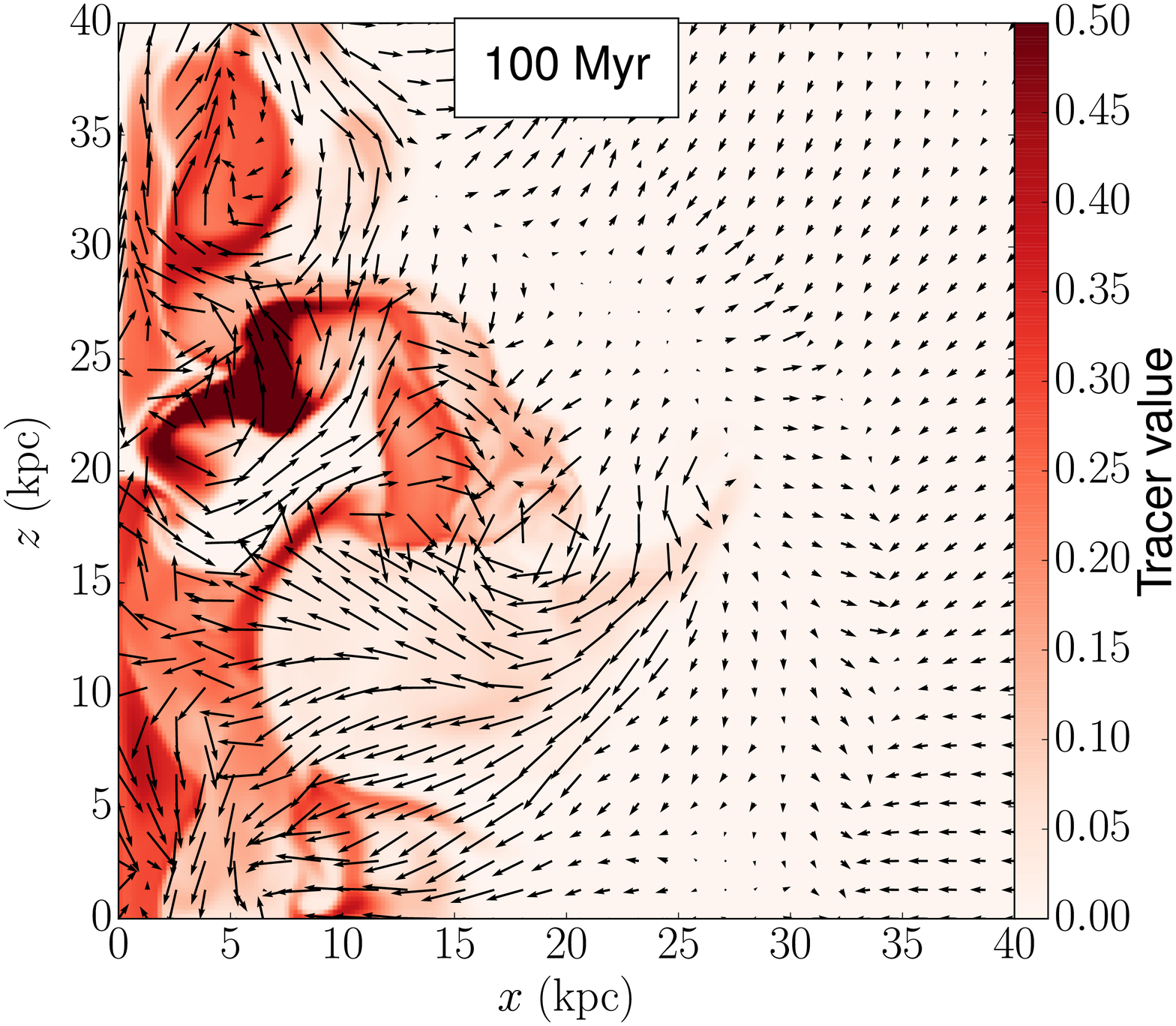}}
\caption{Evolution with time of the gas that at $t=0$ was
contained in a torus whose axis is the $z$-axis and whose cross
section is a circle centered at $(x,z)=(10,5)\kpc$ with a radius
of $r=2.5\kpc$. Shown is the concentration of this gas, as
followed by its tracer, in the meridional plane $y=0$ at four
times. Flow parameters are as in previous figures. The largest
velocity vector corresponds to $v_{\rm m}=400 \kms \simeq 0.5
{\mathcal{M}_{\rm ICM}}$, where ${\mathcal{M}_{\rm ICM}}$ is the
Mach number in the ICM. Higher velocities are marked with arrows with the same length as
that of $v_{\rm m}$.}
 \label{figure: tracers tr4}
\end{figure}

\begin{figure}[!htb]
\centering
\subfigure{\includegraphics[width=0.45\textwidth]{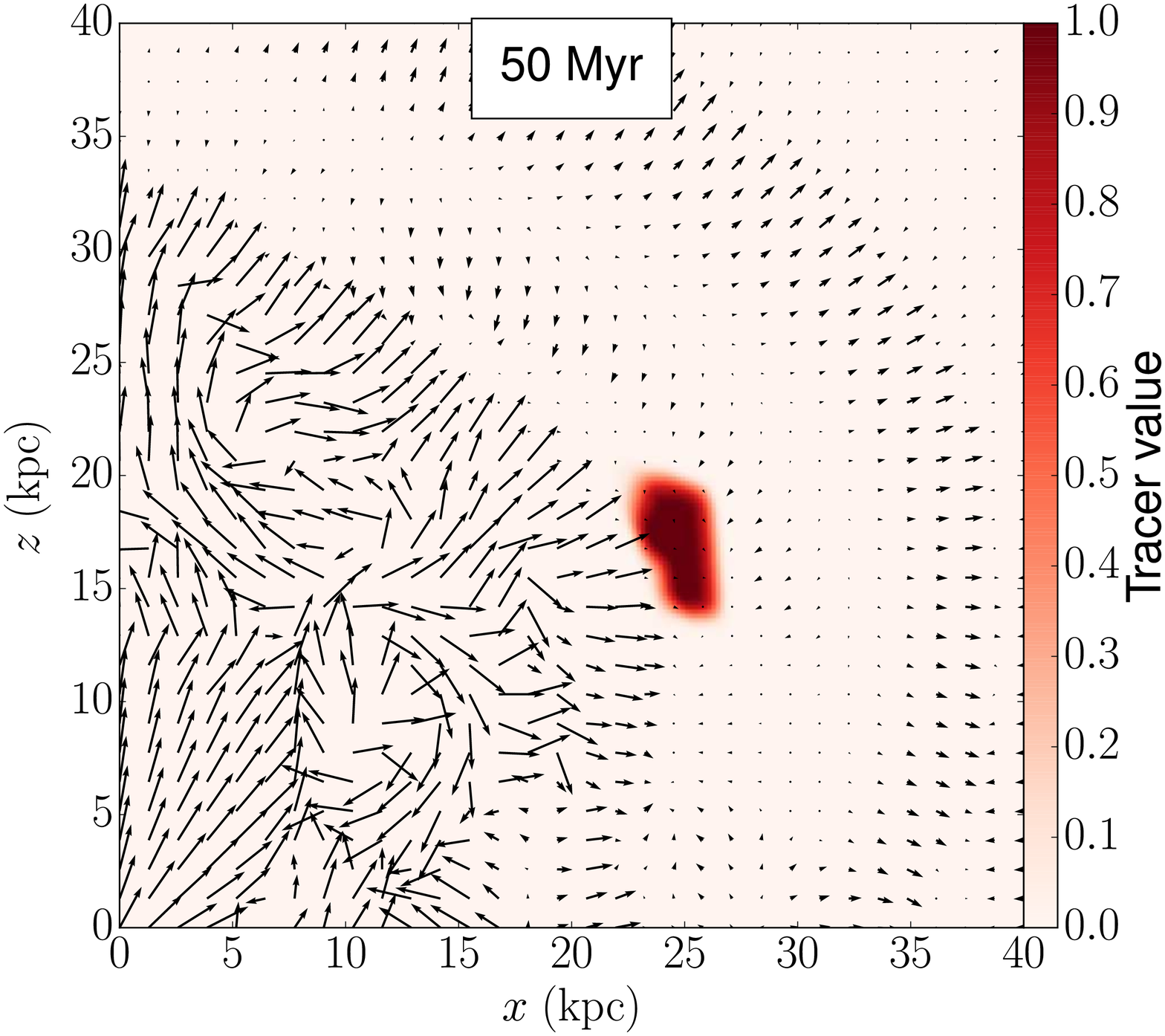}}
\subfigure{\includegraphics[width=0.45\textwidth]{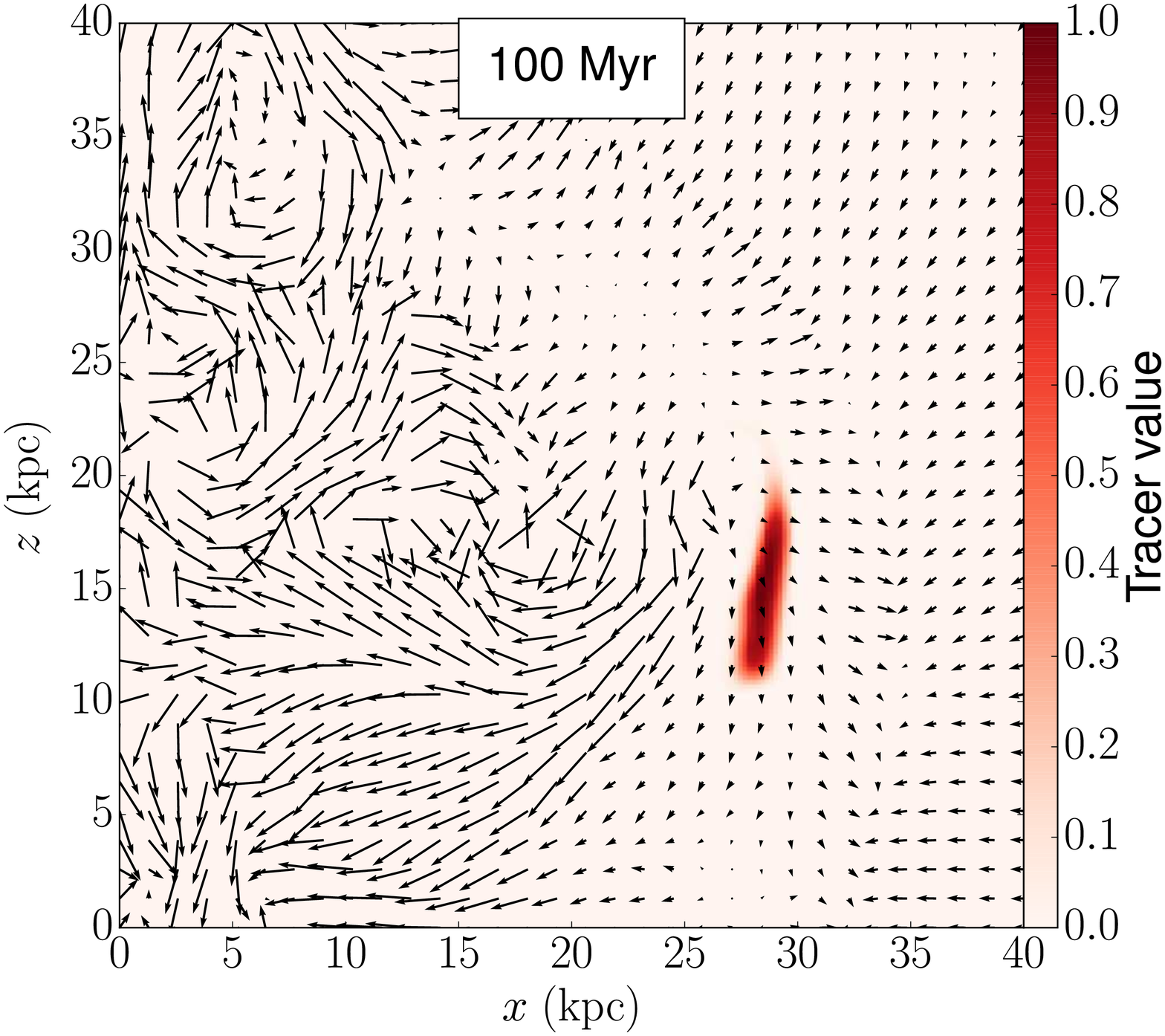}}\\
\subfigure{\includegraphics[width=0.45\textwidth]{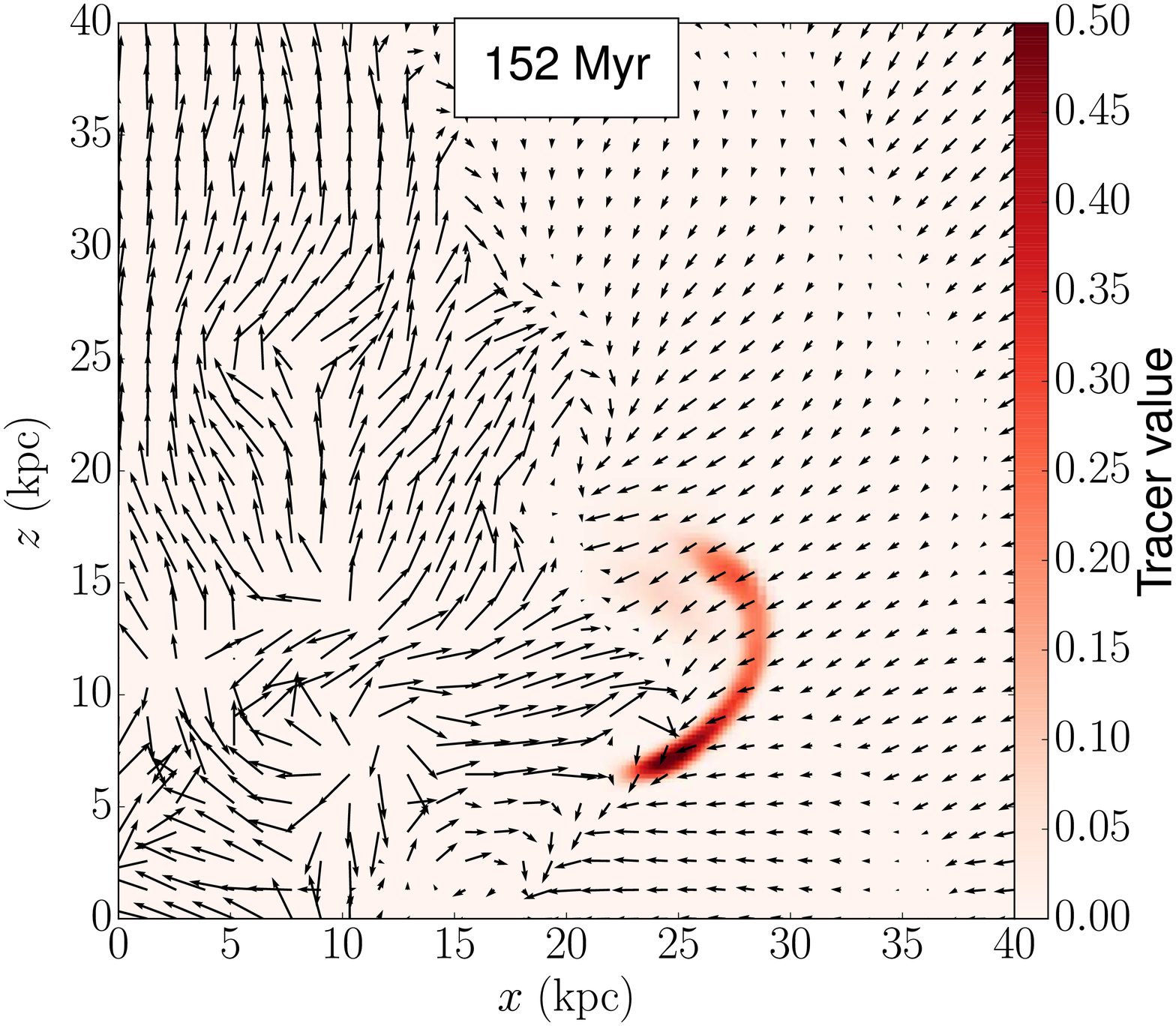}}
\subfigure{\includegraphics[width=0.45\textwidth]{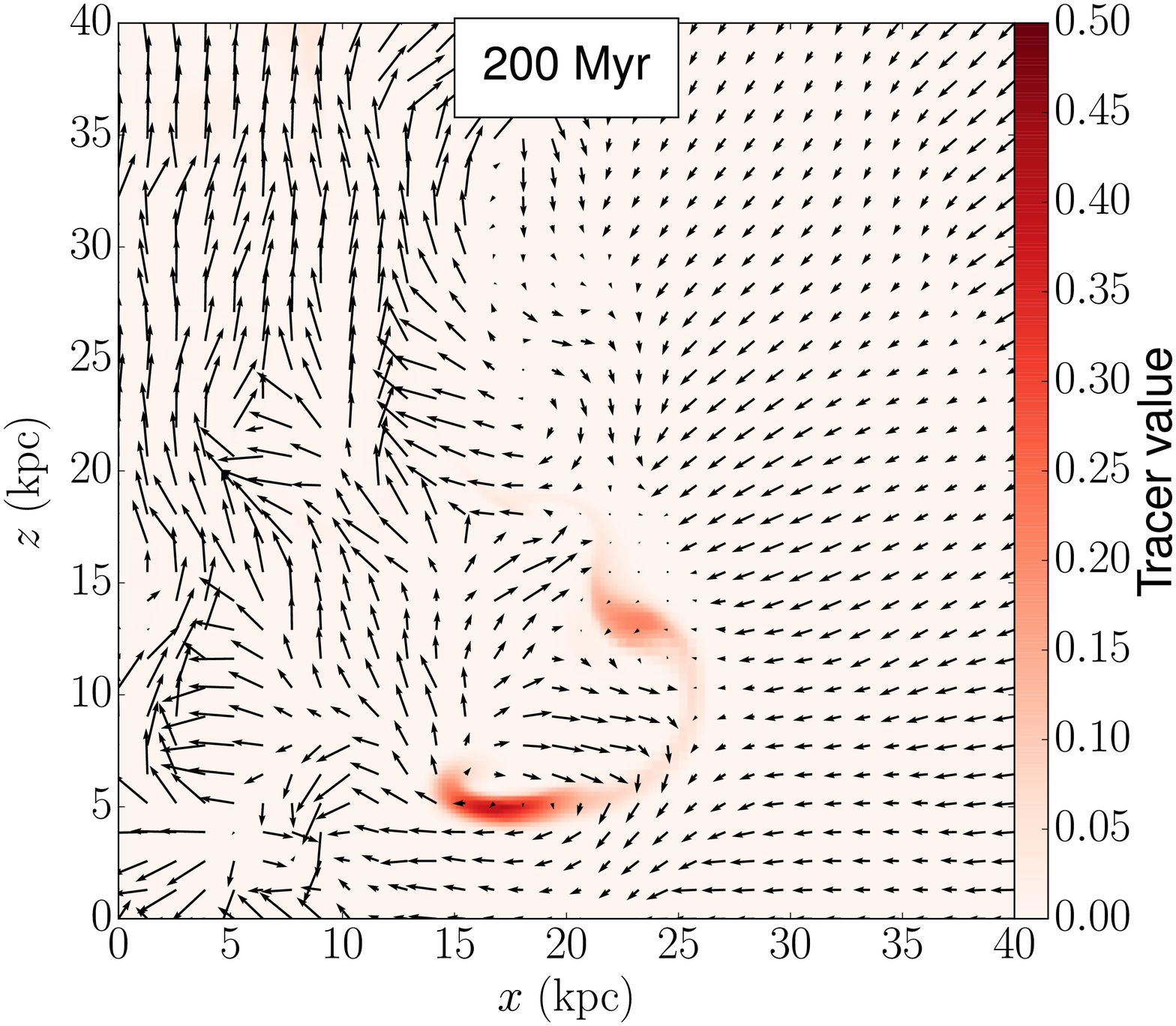}}\\
\subfigure{\includegraphics[width=0.45\textwidth]{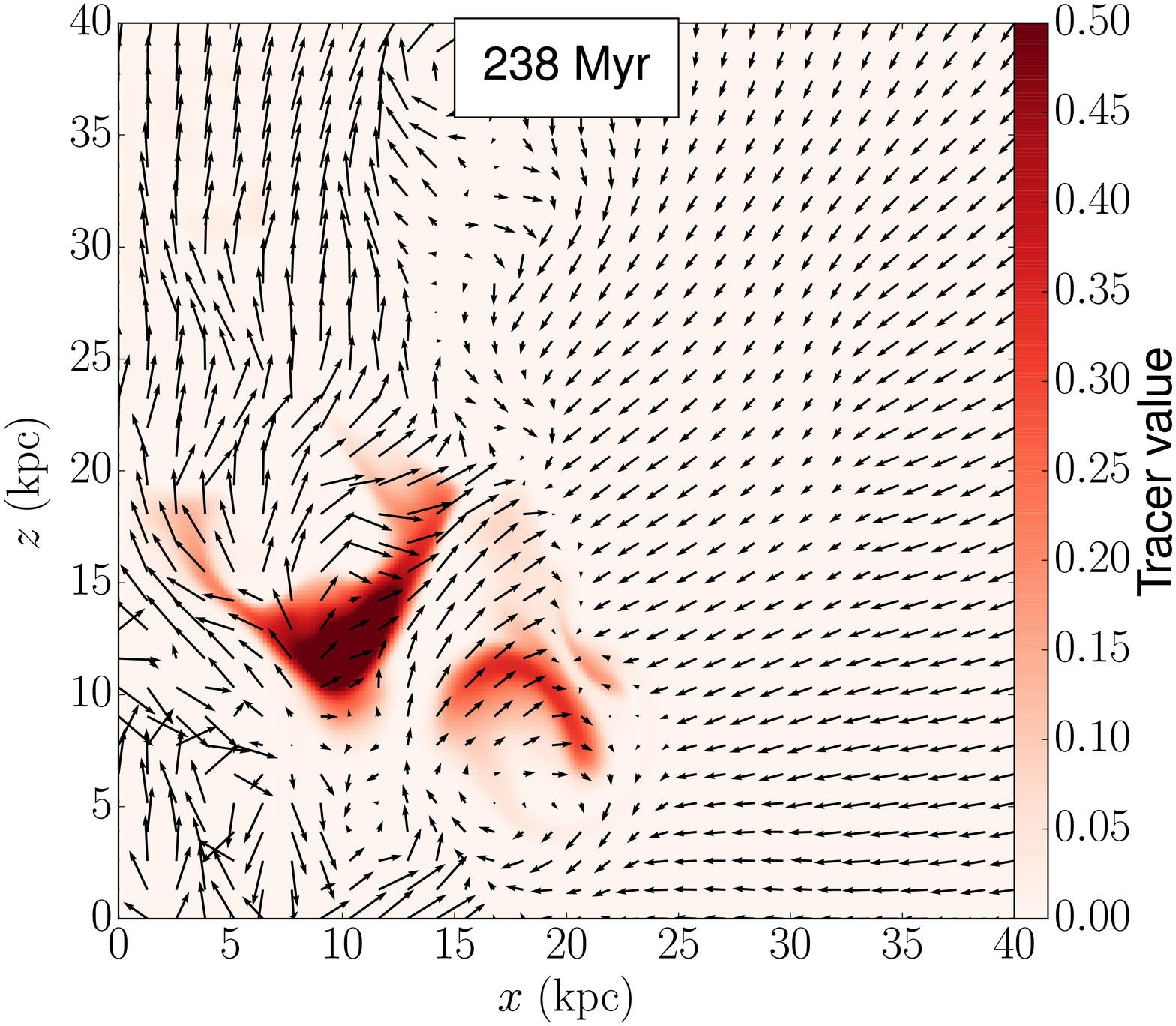}}
\subfigure{\includegraphics[width=0.45\textwidth]{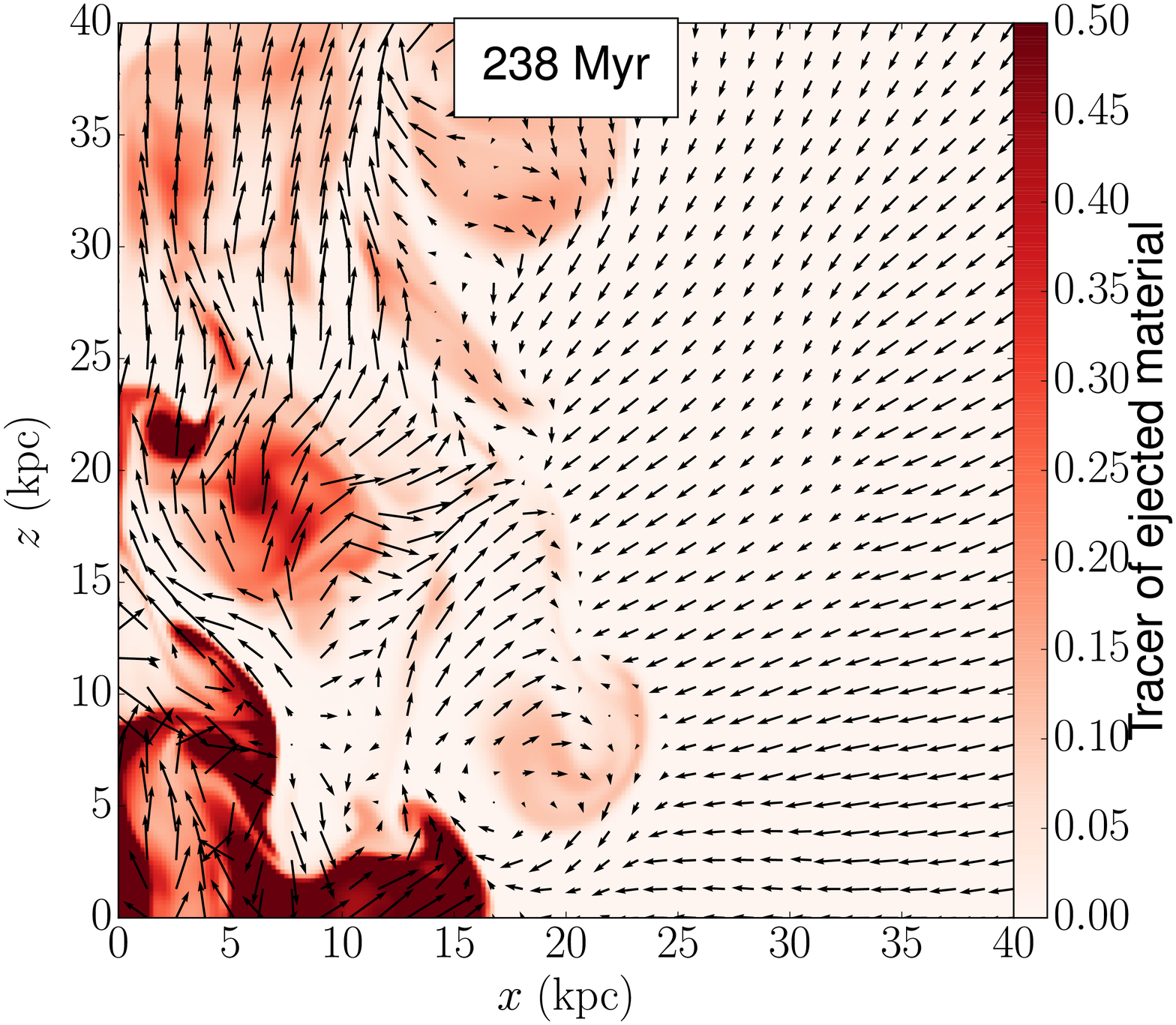}}
\caption{Evolution with time of the gas that at $t=0$ was
contained in a torus whose axis is the $z$-axis and whose cross
section is a circle centered at $(x,z)=(20,15)\kpc$ with a radius
of $r=2.5\kpc$. In the first five panels the concentration of this
gas, as followed by its tracer, is shown in the meridional plane
$y=0$ at five times. The lower-right panel shows the tracer of the
jet. Flow parameters as in previous figures. The largest velocity
vector corresponds to $v_{\rm m}=400 \kms$, a Mach number of about
$0.5$. Higher velocities are marked with arrows with the same length as that of $v_{\rm m}$. }
 \label{figure: tracers tr3}
\end{figure}

In Figs.~\ref{figure: tracers tr4} and \ref{figure: tracers tr3}
we show the evolution of tracers that are frozen-in to the ICM. In
both cases the tracer follows the gas that started in a torus
around the $z$ axis, and the radius of the cross section of the
torus is $2.5 \kpc$. In Figs.~\ref{figure: tracers tr4} we follow
the tracers whose torus cross section is centered at $(x,z)=(10,5)
\kpc$, and in Fig.~\ref{figure: tracers tr3} whose torus cross
section is centered at $(x,z)=(20,15)\kpc$. The bottom right panel
of Fig.~\ref{figure: tracers tr3} shows the tracer of the jet
rather than the tracer of the ICM. Both figures clearly show the
efficient mixing of the ICM with the hot bubble gas.

The tracer in Fig.~\ref{figure: tracers tr4} is initially pushed
outward by the outgoing shock waves and the growing bubble of hot
gas that induces sound waves. However, at $t
\approx 40 \Myr$ it begins to be mixed with hot jet gas, as the
large vortex induced by bubble inflation tears it apart. At $t
\approx 100-150 \Myr$ it becomes almost completely mixed with its
surroundings as well as with the jet gas.

The tracer of the gas starting further out and presented in
Fig.~\ref{figure: tracers tr3} follows a similar pattern, but
begins to be mixed with the hot bubble gas at much later times.
Even that it is pushed out to $\approx 30 \kpc$, it eventually
flows inward and mixes with the hot bubble gas and heats up, as we
show in section \ref{s-heating-the-icm}. In the lower-right panel
of Fig.~\ref{figure: tracers tr3} we show the tracer of the jet.
The two lower panels that are given at the same time clearly
present the efficient mixing of the ICM and the hot bubble gas.
For example, a clear vortex mixing the two media is seen
circulating around $(x,z)=(17,5) \kpc$.

An inflow of gas near the equatorial plane is seen at the latest
time in Fig.~\ref{figure: tracers tr3}. This carries ICM gas toward the jets, followed by mixing and
heating. This demonstrates again that the jets can heat material
even in regions further out and near the equatorial plane. If the
jets continue to be active for a very long time along the same
axis, a massive meridional flow will develop in the ICM
\citep{Sokeretal2015}.
This flow carries cooler ICM gas towards the center and mixes it with the hot bubble gas.
This is an indirect mechanism by which a jet may heat up gas in regions further away from its axis.

\section{HEATING THE ICM}
\label{s-heating-the-icm}

We turn to present the energy history of the ICM. In order to follow specific
regions we mark them with `tracers' as explained in section
\ref{s-flow-structure}. The total mass of a traced gas is given by
the sum $M_{\rm tr}=\Sigma \xi_i M_i$ over all grid cells. As the
tracers are advected with their associated mass, the total traced
mass is constant with time, as long as traced gas does not leave
the grid. Since there are outflow boundary conditions on the grid
faces $x=50 \kpc$, $y=50 \kpc$ and $z=50 \kpc$, and traced gas is
dragged by the jets after some time, traced gas does eventually
leave the grid.  When that happens we stop following the thermal
evolution of that traced gas.

\begin{figure}[!htb]
\centering
\subfigure{\includegraphics[width=0.32\textwidth]{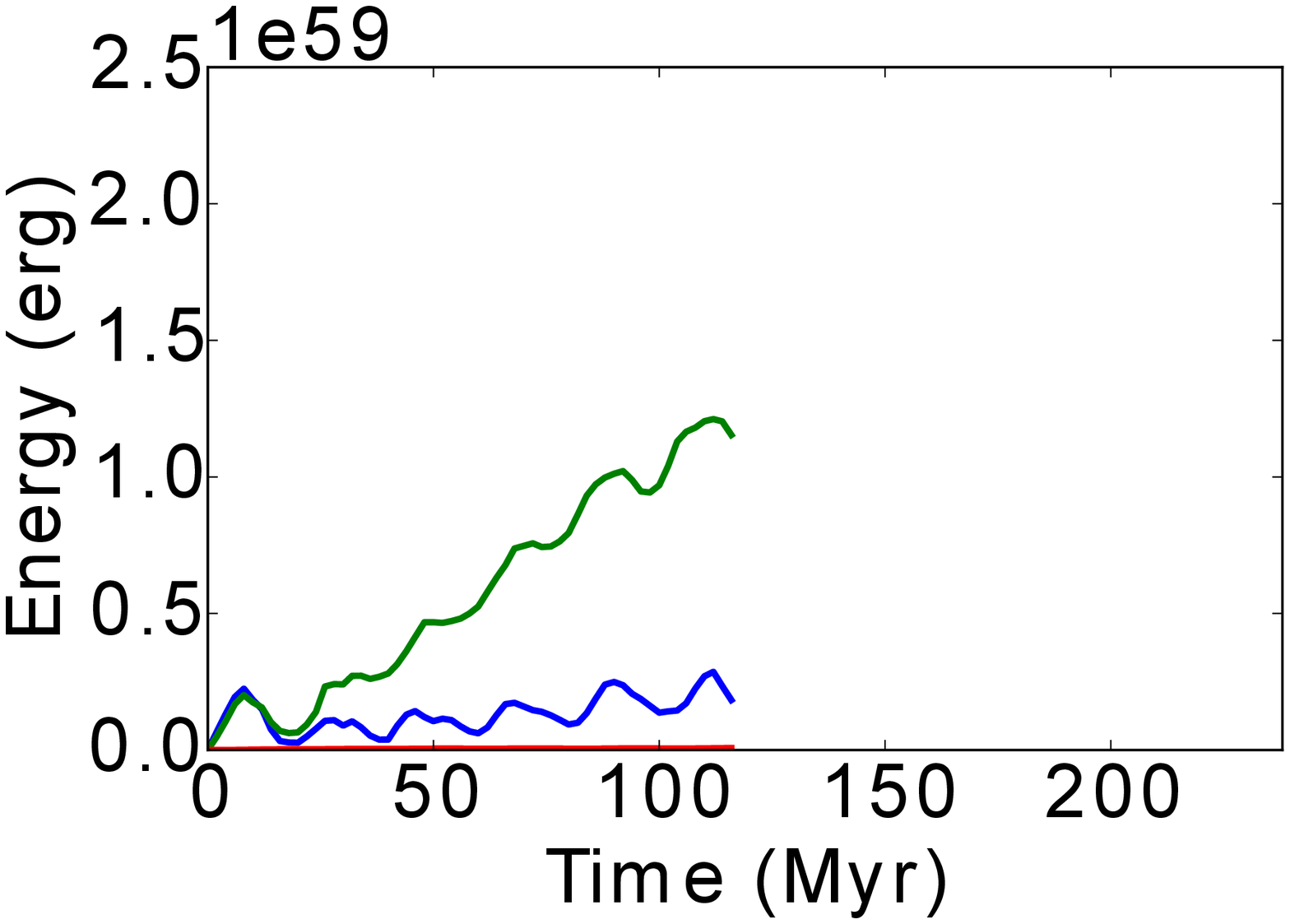}}
\subfigure{\includegraphics[width=0.32\textwidth]{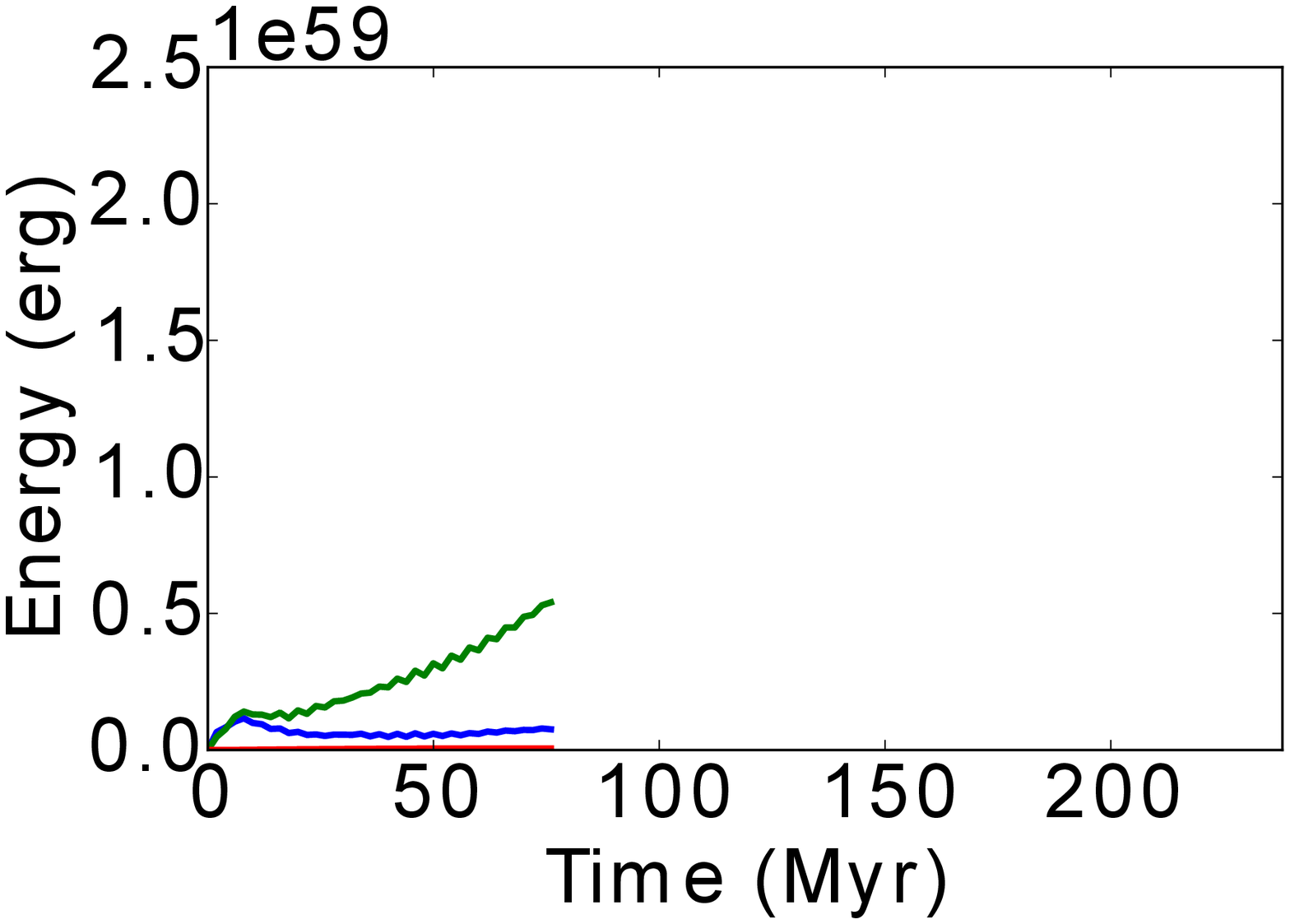}}
\subfigure{\includegraphics[width=0.32\textwidth]{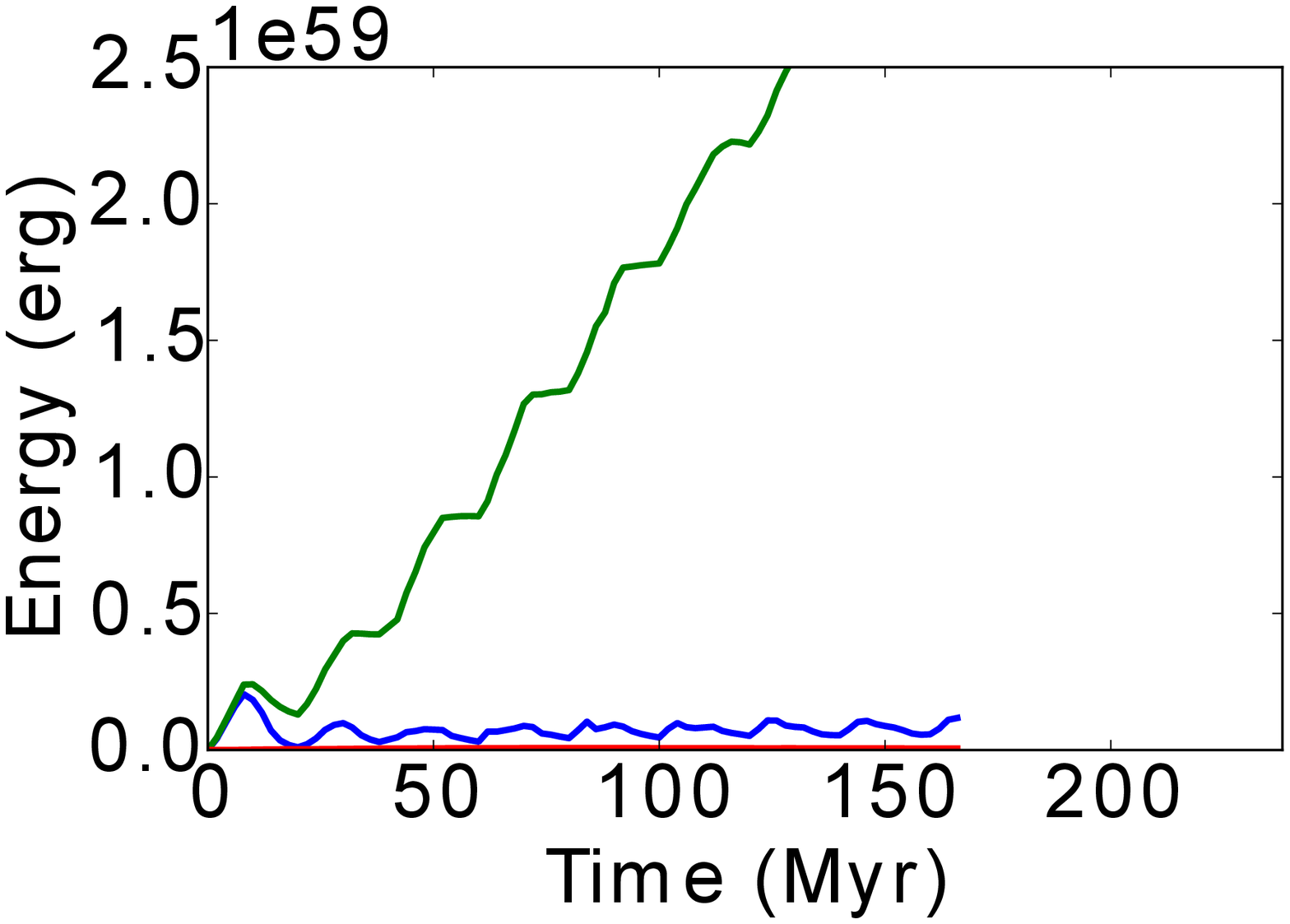}}\\
\subfigure{\includegraphics[width=0.32\textwidth]{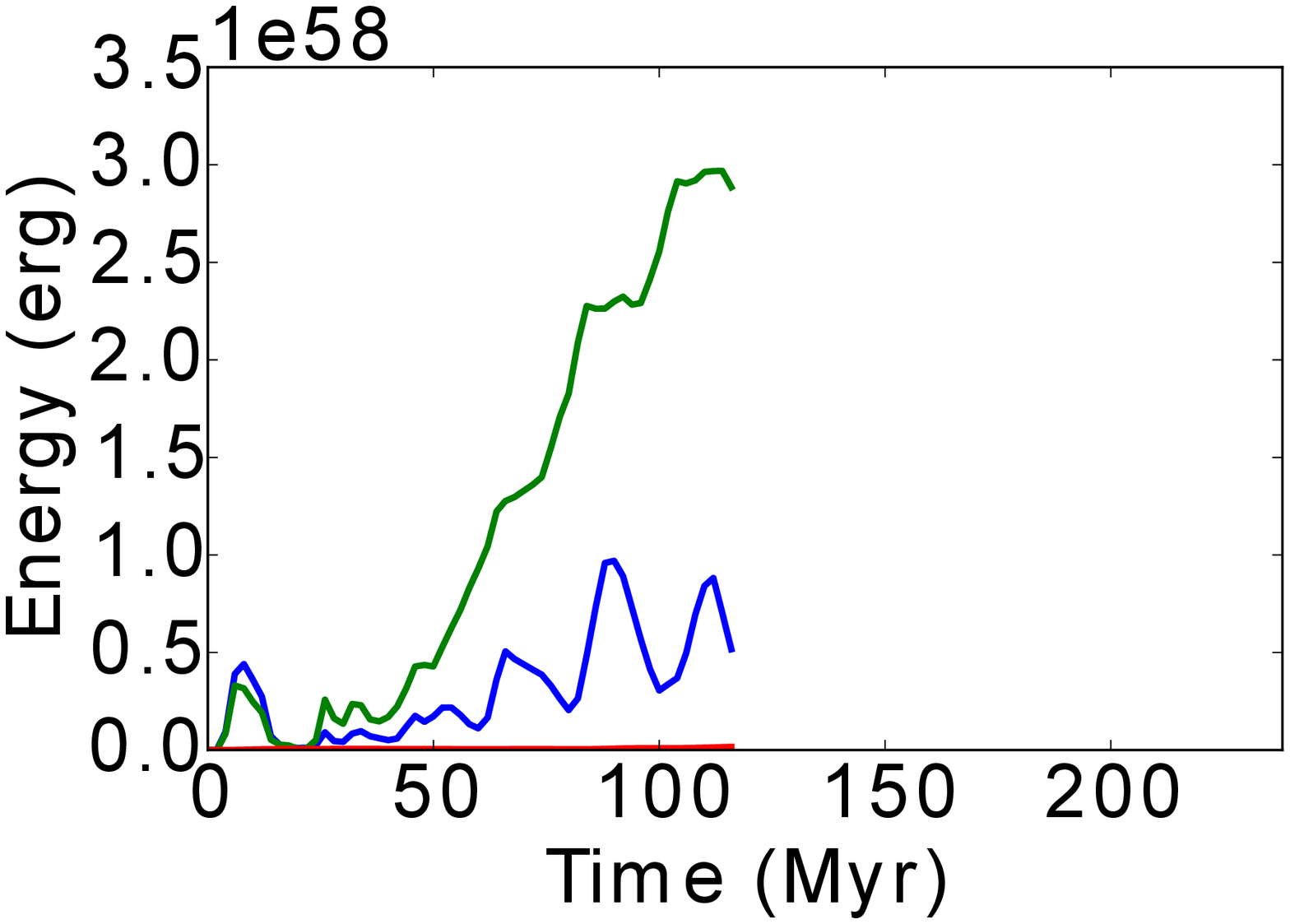}}
\subfigure{\includegraphics[width=0.32\textwidth]{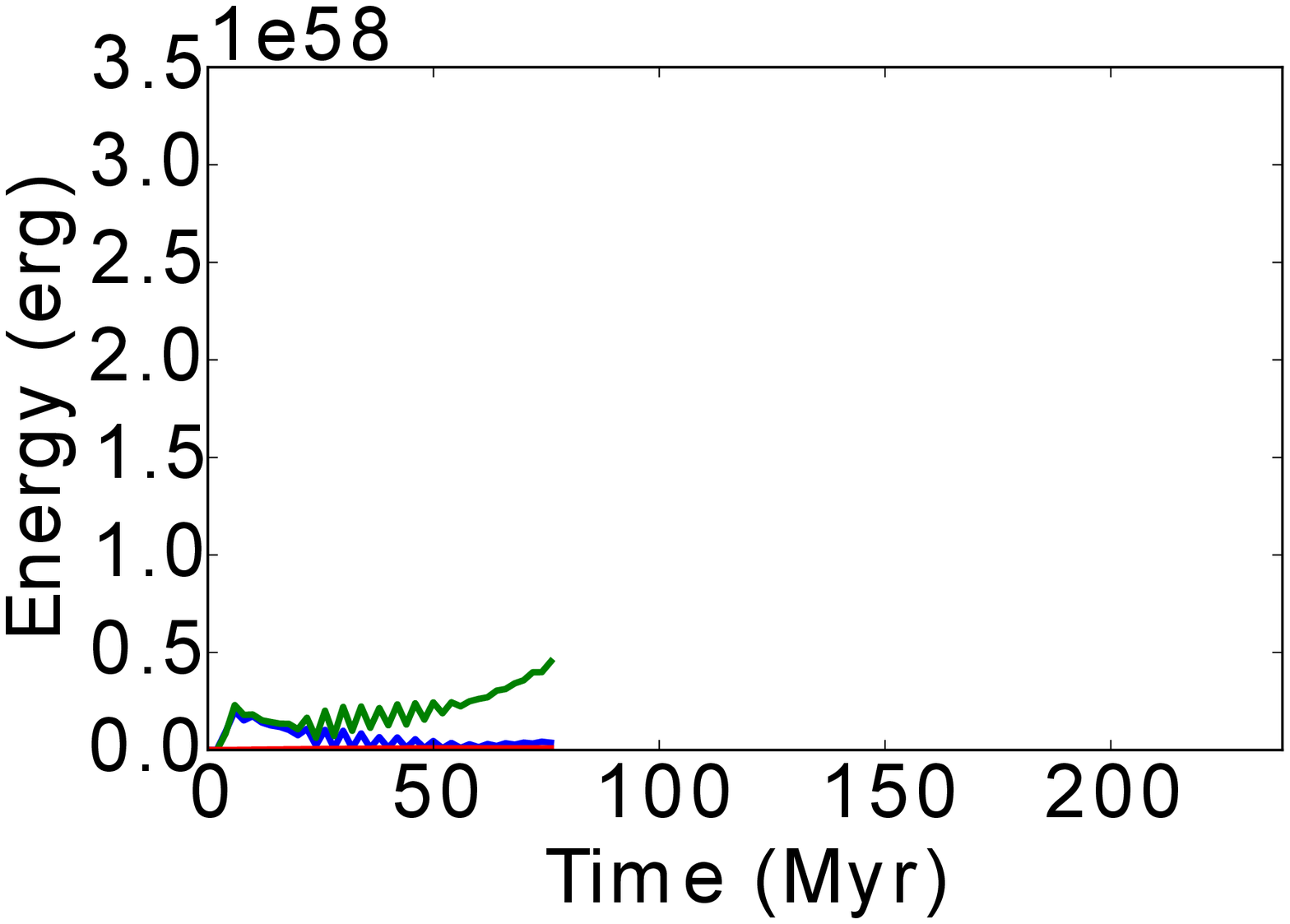}}
\subfigure{\includegraphics[width=0.32\textwidth]{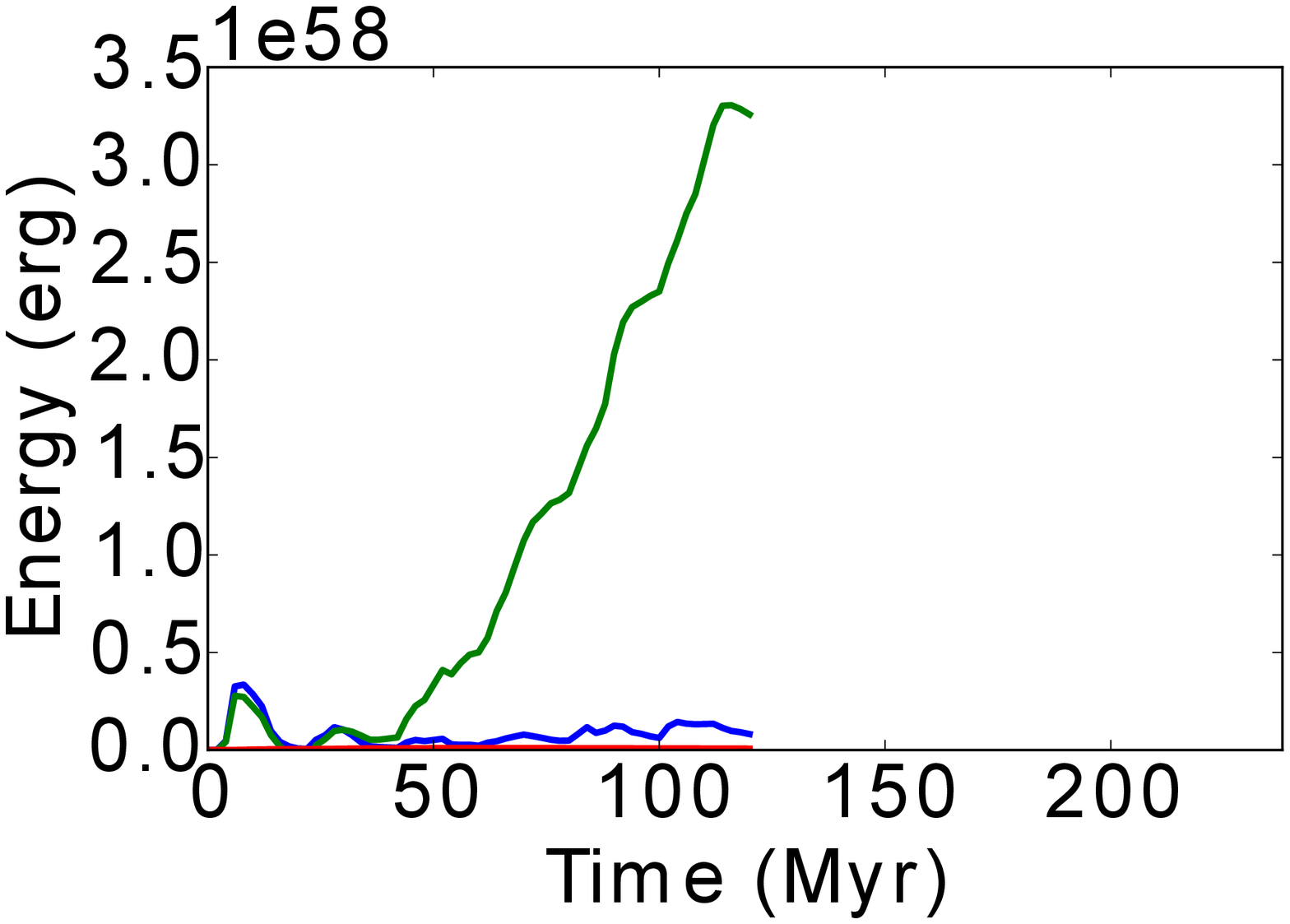}}\\
\subfigure{\includegraphics[width=0.32\textwidth]{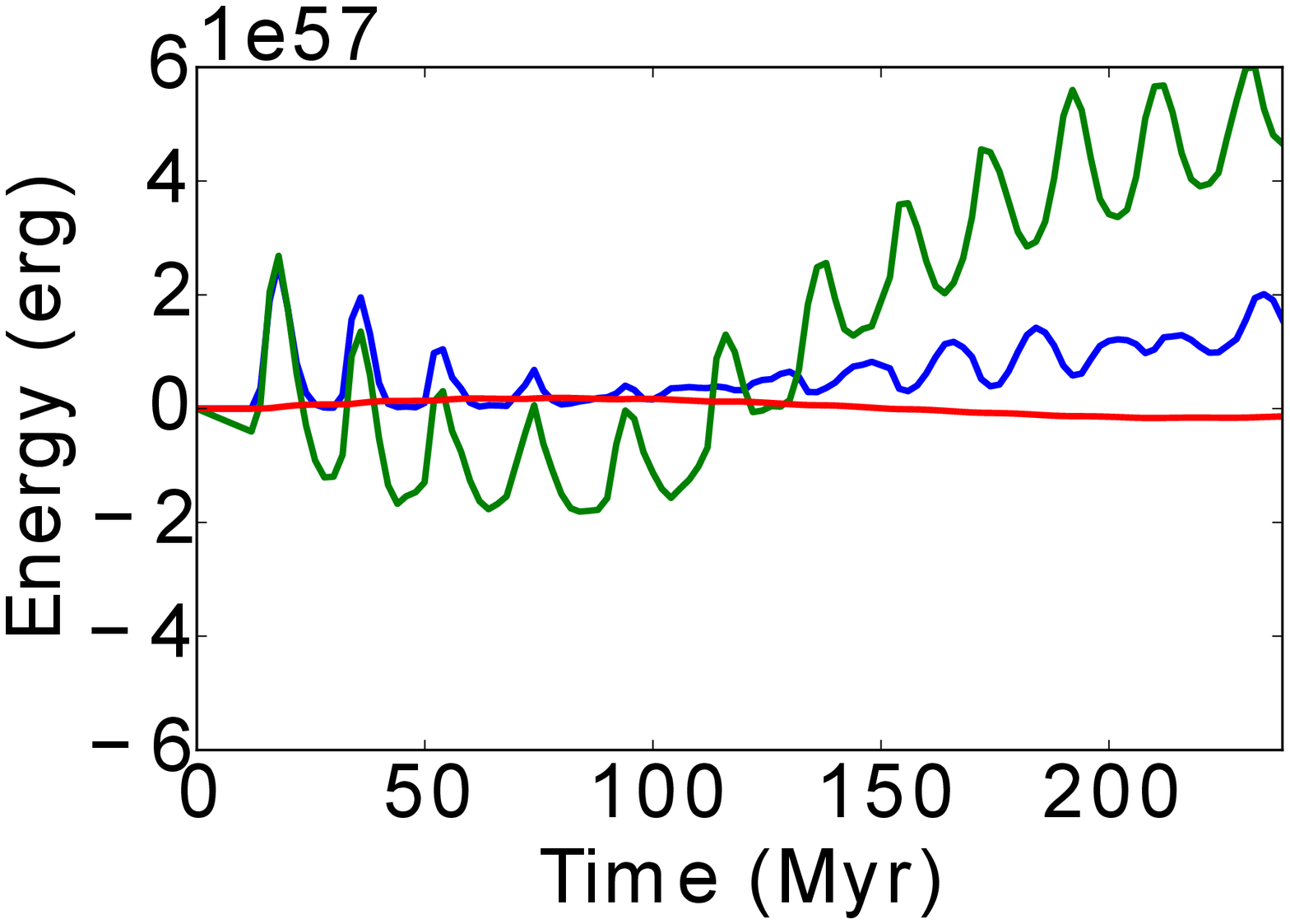}}
\subfigure{\includegraphics[width=0.32\textwidth]{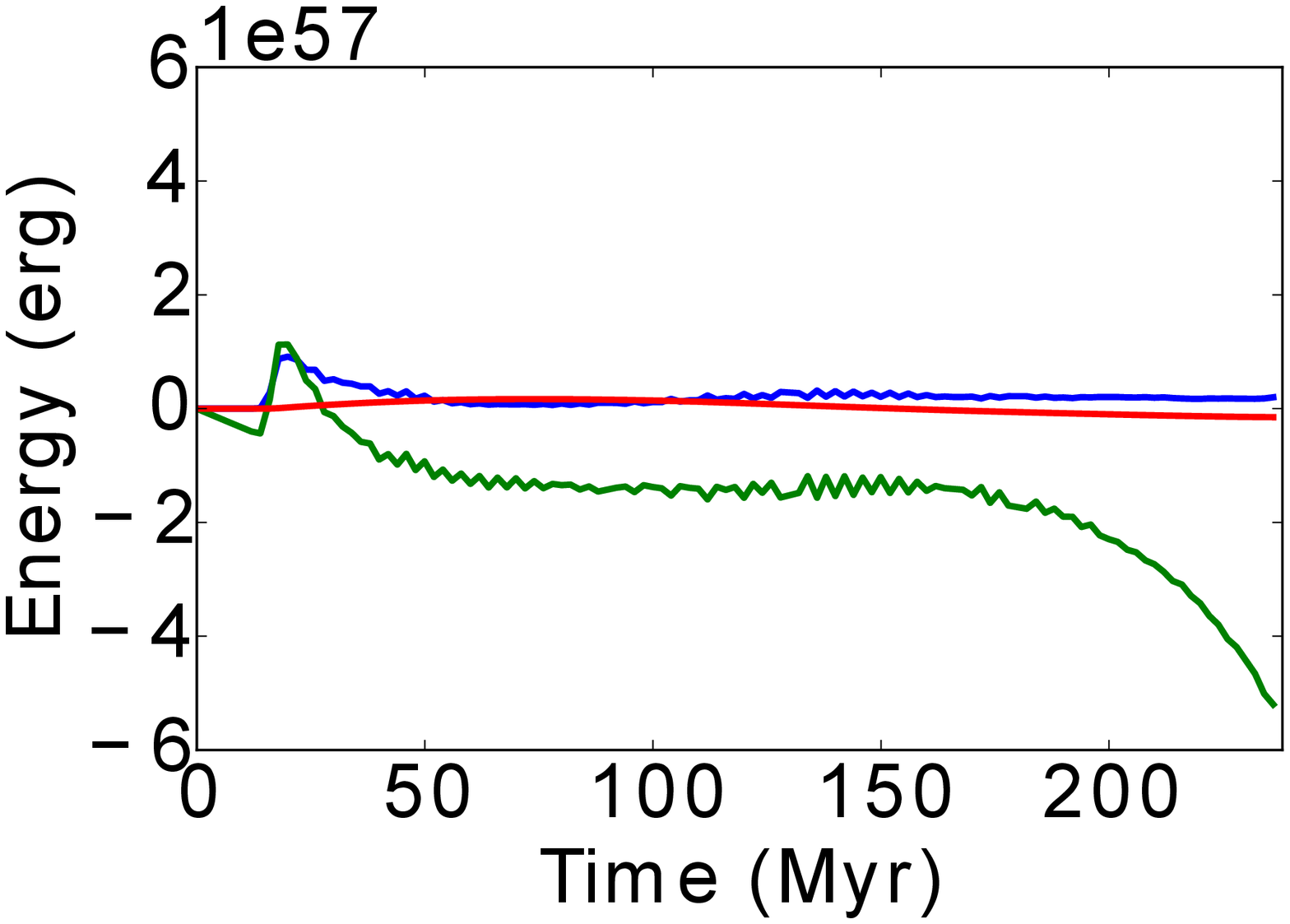}}
\subfigure{\includegraphics[width=0.32\textwidth]{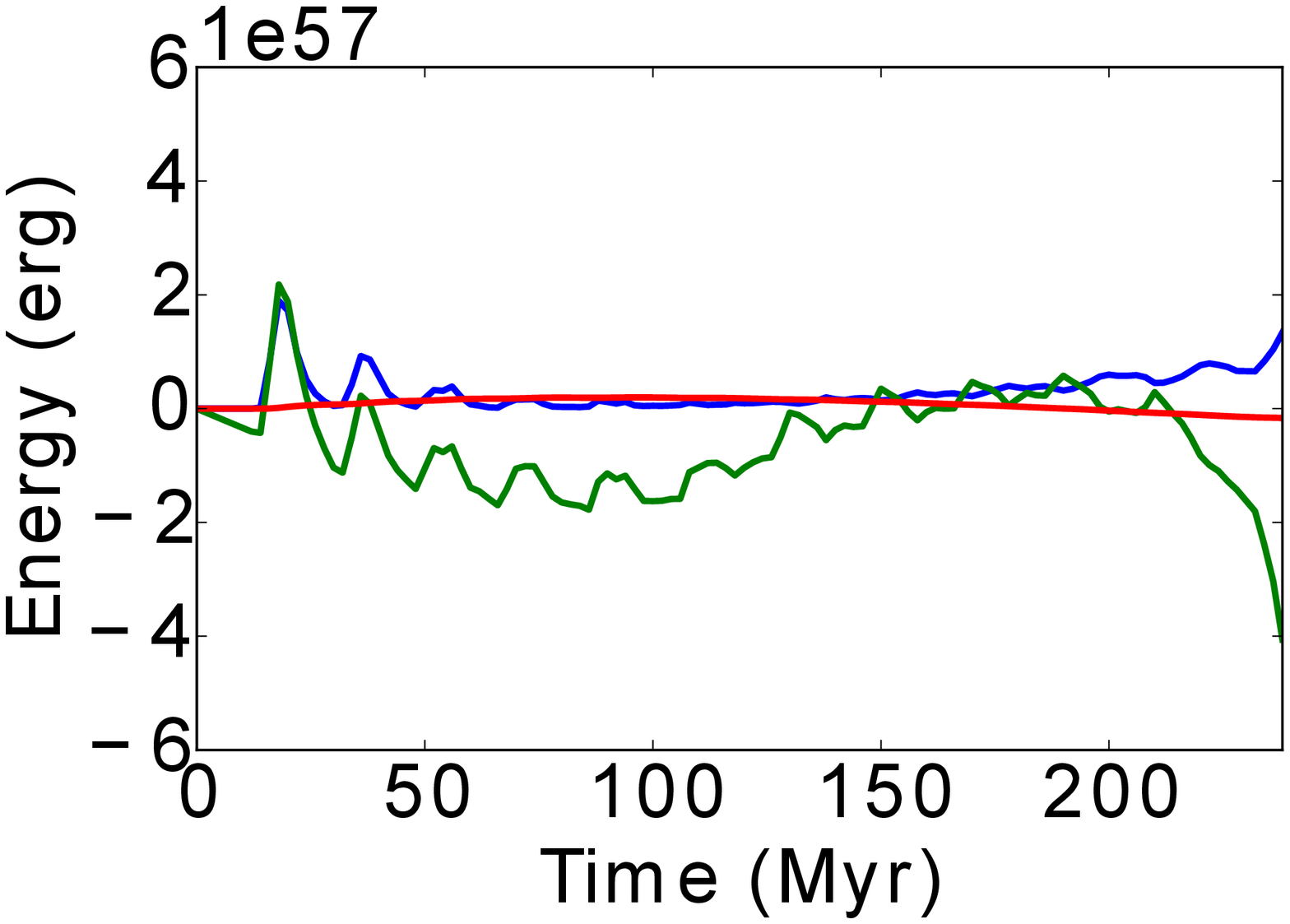}}
\caption{
The energy history of different traced regions of the
nominal simulation studied here as well as in two additional simulations.
The three panels in the left-hand column represent the nominal simulation.
In the second simulation, Run B, presented in the middle column, the jet is periodically turned on for $2\Myr$ and off for
$2\Myr$. The rest of the parameters are as in the nominal simulation used in the rest of this work.
In the third simulation, Run C, presented in the right-hand column, the jet mass deposition rate is decreased by a factor of $10$
and the jet velocity is increased by a factor of $\sqrt{10}$ with respect to the nominal simulation,
such that the jet's power is unchanged.
Otherwise, the setup is the same.
In the three panels in the top row we present the
evolution with time of three energy components of the ICM gas that
starts (before the jets become active) inside an eighth of
ball with $r=15\kpc$ centered at the origin. It is an eighth of a
ball as we simulate one eighth of the space. The green line in each panel represents the internal
energy, the blue line represents the kinetic energy, and
the red line represents the gravitational energy of this
traced gas. The middle-row panels show the energy histories of the
torus shown in Fig.~\ref{figure: tracers tr4}, and the panels in the bottom row
show the energy histories of the torus shown in Fig.~\ref{figure: tracers tr3}.
All energies are shown relative to their values at
$t=0$. The initial internal energies $E_{\rm in}(0)$ of the traced regions, from top to bottom,
are $3.1 \times 10^{58} \erg$, $5.5 \times 10^{57} \erg$ and  $1.1 \times 10^{58} \erg$, respectively.
The top- and middle-row panels are cut off at the time when the
traced material starts leaving the grid.
}
\label{figure: energy history}
\end{figure}

We use the tracers to define the total energy $E_{\rm
tr}$ of a traced region as
\begin{equation}
E_{\rm tr} \equiv
 \Sigma_i \xi_i E_i,
\end{equation}
where $i$ runs over all grid cells. In this work, $E_{\rm tr}$
stands for the traced gas's total kinetic energy, internal energy, or gravitational
energy, and $E_i$ stands for the corresponding energy in cell $i$. In Fig.~\ref{figure: energy history} we
present the energy histories of different traced regions in the
simulations. The internal energy and the gravitational energy are
drawn with respect to their initial value which is different than
zero. The initial kinetic energy is zero. The initial thermal
energy of each traced region is given in the caption.

In addition to the nominal jet parameters specified in sections \ref{s-numerical-setup} and \ref{s-flow-structure},
we conducted two simulations with varying parameters, in order to check the robustness of our conclusions.
The nominal simulation is represented in the three panels in the left-hand column of Fig.~\ref{figure: energy history}.
Two additional simulations, Run B and Run C, are represented in the middle and right-hand columns respectively.
We shall elaborate on them below.

Consider first the results of the nominal simulation, i.e., the left-hand column.
In the panels in the top row of Fig.~\ref{figure: energy history} we present
the energy evolution of mass located at $t=0$ inside a sphere of
radius $r = 15 \kpc$ centered at the origin. Because of the structure
of our grid, only an eighth of the sphere is traced. The initial
internal energy of this traced gas in the simulated eighth is $E_{\rm in}(0)= 3.1 \times 10^{58} \erg$,
such that the maximum value of the extra thermal energy in the
plot is $\Delta=[E_{\rm in}(\rm max) - E_{\rm in}(0)]/E_{\rm
in}(0) = 390 \%$ (in the nominal case) above the initial internal energy. The calculation
is stopped when traced gas starts leaving the computational grid.

The panels in the middle row present the energy history of the traced gas
whose spreading history is displayed in Fig.~\ref{figure: tracers
tr4}. In this case $E_{\rm in}(0)= 5.5 \times 10^{57} \erg$, and $\Delta=[E_{\rm
in}(\rm max) - E_{\rm in}(0)]/E_{\rm in}(0) = 550 \%$ (in the nominal case). The
calculation is stopped when traced gas starts leaving the
computational grid. The panels in the bottom row present the energy history of
the traced mass whose spreading history is shown in
Fig.~\ref{figure: tracers tr3}. In this case $E_{\rm in}(0)= 1.1 \times 10^{58}
\erg$, and $\Delta=[E_{\rm in}(\rm max) - E_{\rm in}(0)]/E_{\rm
in}(0) = 50 \%$ (in the nominal case).

A general characteristic behavior emerges from the energy
evolution of the traced masses presented here, as well as of other
traced regions we have studied but left out of this text. The
variation on a time scale of $20 \Myr$ corresponds to the time
period of the jet activity. Each active phase lasts for $10 \Myr$,
followed by a quiescence phase of $10 \Myr$. The periodic jet
activity induces gas motion in the ICM such that the kinetic
energy of the ICM at each point varies with time. This variation
is more or less in a periodic manner, with only a moderate
increase in the kinetic energy over several periods. The
gravitational energy changes by a negligible amount compared to
both the kinetic and internal energies. The main energy increase
over several periods comes from an increase of the internal
energy.

Consider the middle panel in the left-hand column of Fig.~\ref{figure: energy history},
which shows the energy history of the tracer presented in
Fig.~\ref{figure: tracers tr4}. There are some variations due to
two shock waves that pass through the traced region. The shocks
increase the internal energy, but the gas re-expands and the net
thermal energy gain is negligible. This traced mass begins to mix
with hot bubble gas at $t \approx 40 \Myr$. From that time on the
mean internal energy increases significantly. The upper right
panel of Fig.~\ref{figure: tracers tr4} shows the traced gas at
time $t = 50 \Myr$, while both upper panels of Fig.~\ref{figure: tracers
tr1} show the presence of the jet material at the same time. It
is apparent that the hot shocked jet (bubble) gas is being mixed
vigorously with the ICM gas marked by the tracer. The mixing is
facilitated by the vortices induced by the jet-inflated bubble.

Similar heating by mixing happens in the tracer presented in
Fig.~\ref{figure: tracers tr3}, whose energy history is shown in
the bottom panel in the left-hand column of Fig.~\ref{figure: energy history}. Periodic
heating by shocks is clearly seen, although over a long time
shocks cannot compete with radiative cooling. Indeed, at early
times the internal energy decreases. Here the mixing begins at $t
\sim 110 \Myr$. In the $t=100 \Myr$ panel of Fig.~\ref{figure:
tracers tr3} we can see a vortex that is just about to touch the
traced region and mix it with hot bubble gas that can be seen in
the $t=100 \Myr$ panel of Fig.~\ref{figure: tracers tr1}.

The heating is mainly due to mixing of the ICM gas with shocked
jet material, i.e., the hot bubble gas. Both the mixing that leads
to an increase in internal energy and the turbulent energy are
driven by vortices. It is hard to estimate directly the amount of
turbulent kinetic energy of the ICM from the simulations, because
the grid resolution is not high enough to capture all relevant
scales. However, we give rough arguments that the kinetic energy
we calculate is not much smaller than the corresponding `true'
turbulent energy. Firstly, the kinetic energy is calculated with
respect to the center of mass of the whole system, and not with
respect to a local mean flow of the gas. Thus, it overestimates
the turbulent energy as it is usually defined. Secondly, our
calculated kinetic energy does not properly include the kinetic
energy in smaller scales than the grid resolution, and thus it
underestimates the turbulent kinetic energy. However, the kinetic
energy is mainly due to vortices which are significantly larger
than the grid resolution. Therefore, we do not expect that the
underestimation of the turbulent kinetic energy in small
unresolved scales is significant. Thirdly, some of the kinetic
energy we count is due to sound waves that propagate out and do
not heat the inner region. Finally, motion in smaller scales is
less energetic, and it also tends to dissipate to internal energy.

\begin{figure}[!htb]
\centering
\subfigure{\includegraphics[width=0.32\textwidth]{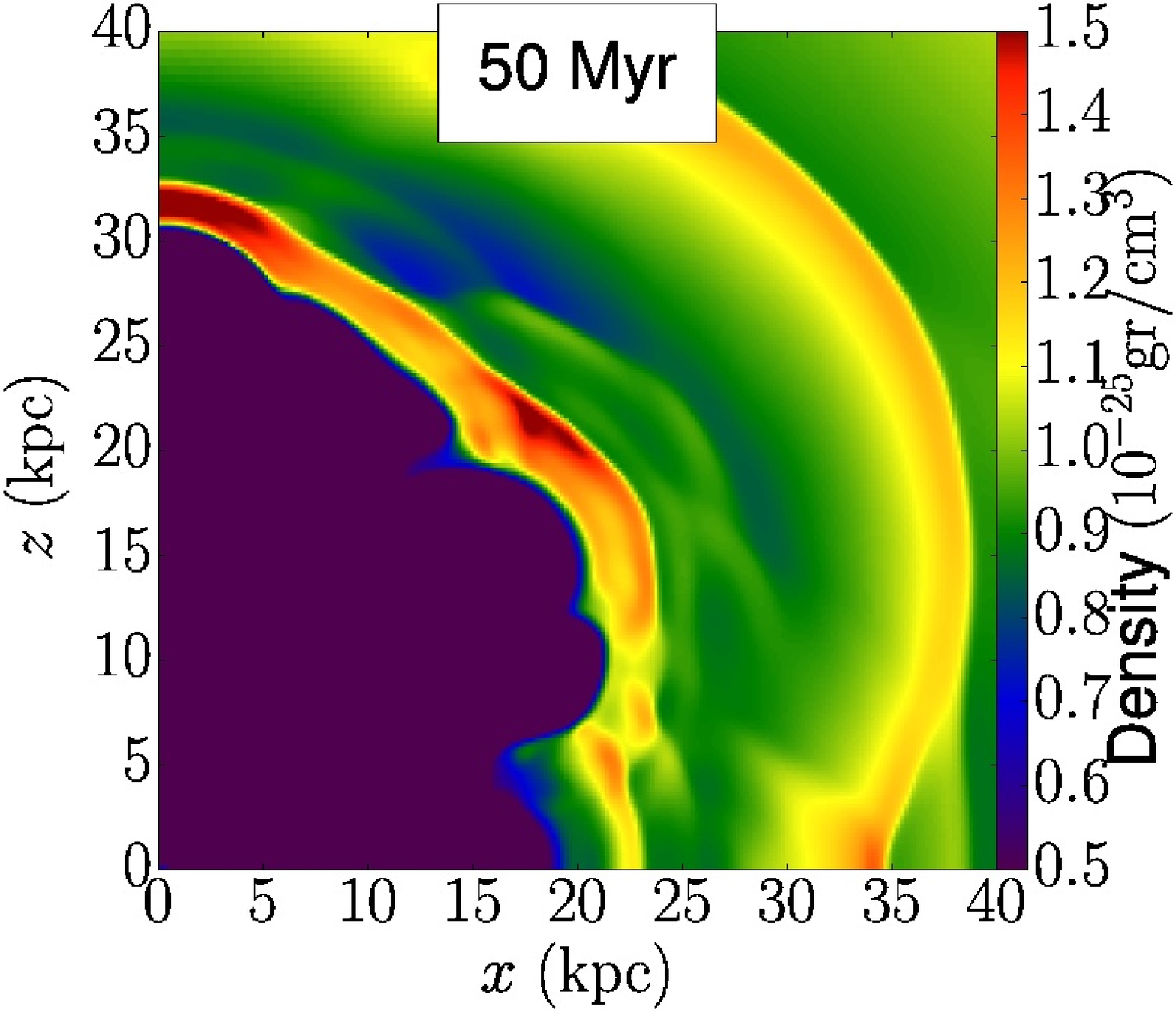}}
\subfigure{\includegraphics[width=0.32\textwidth]{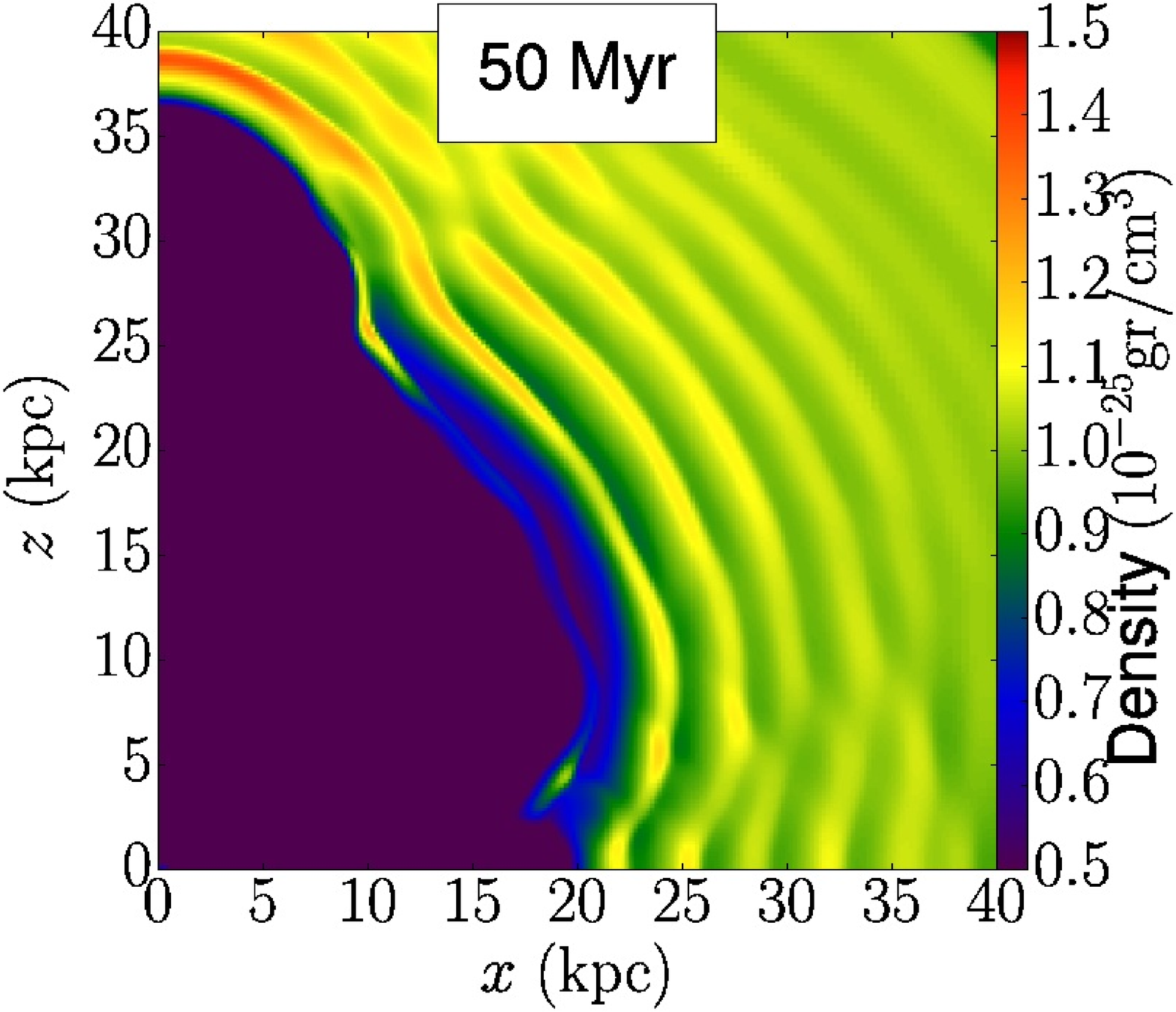}}
\subfigure{\includegraphics[width=0.32\textwidth]{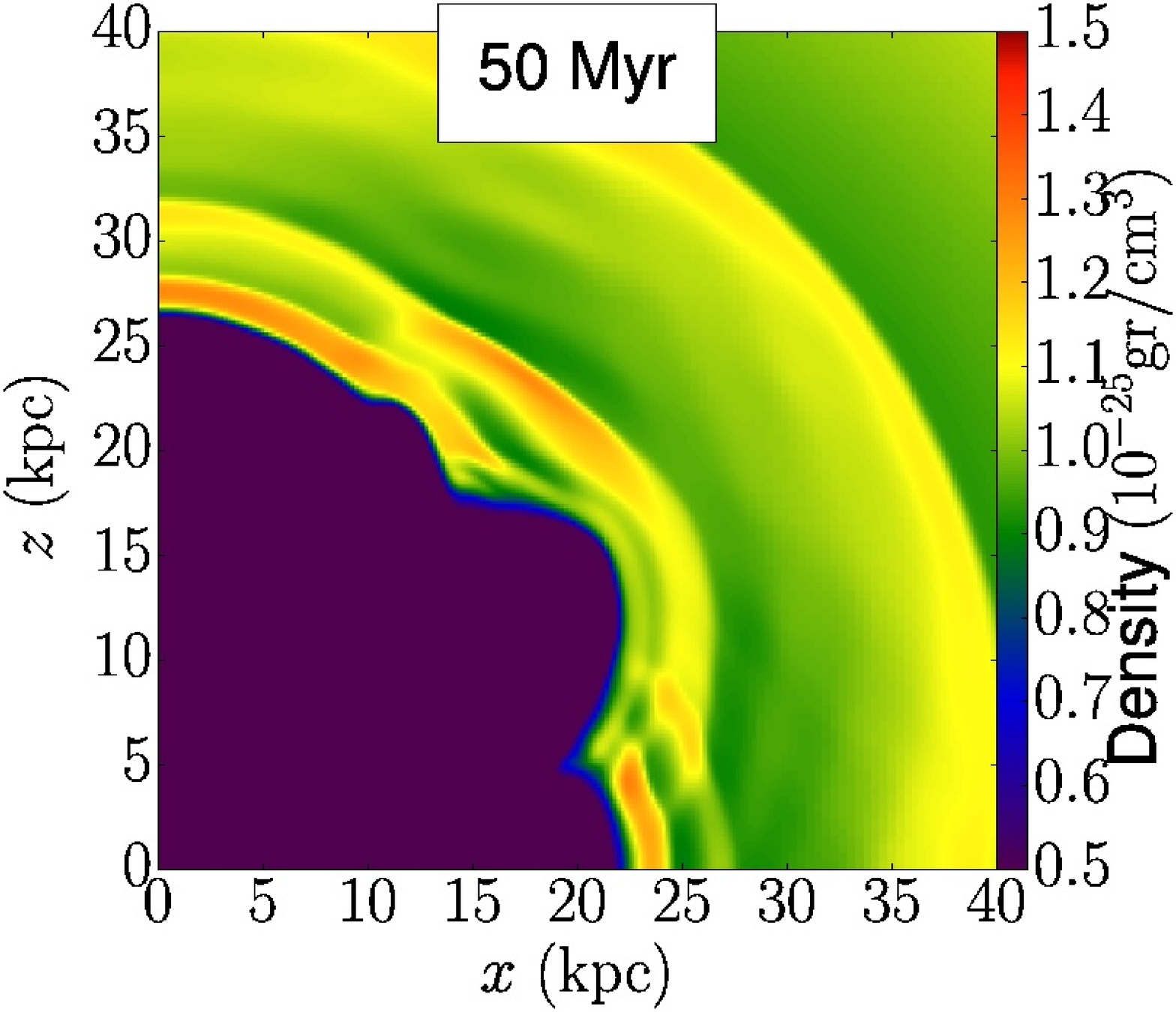}}\\
\subfigure{\includegraphics[width=0.32\textwidth]{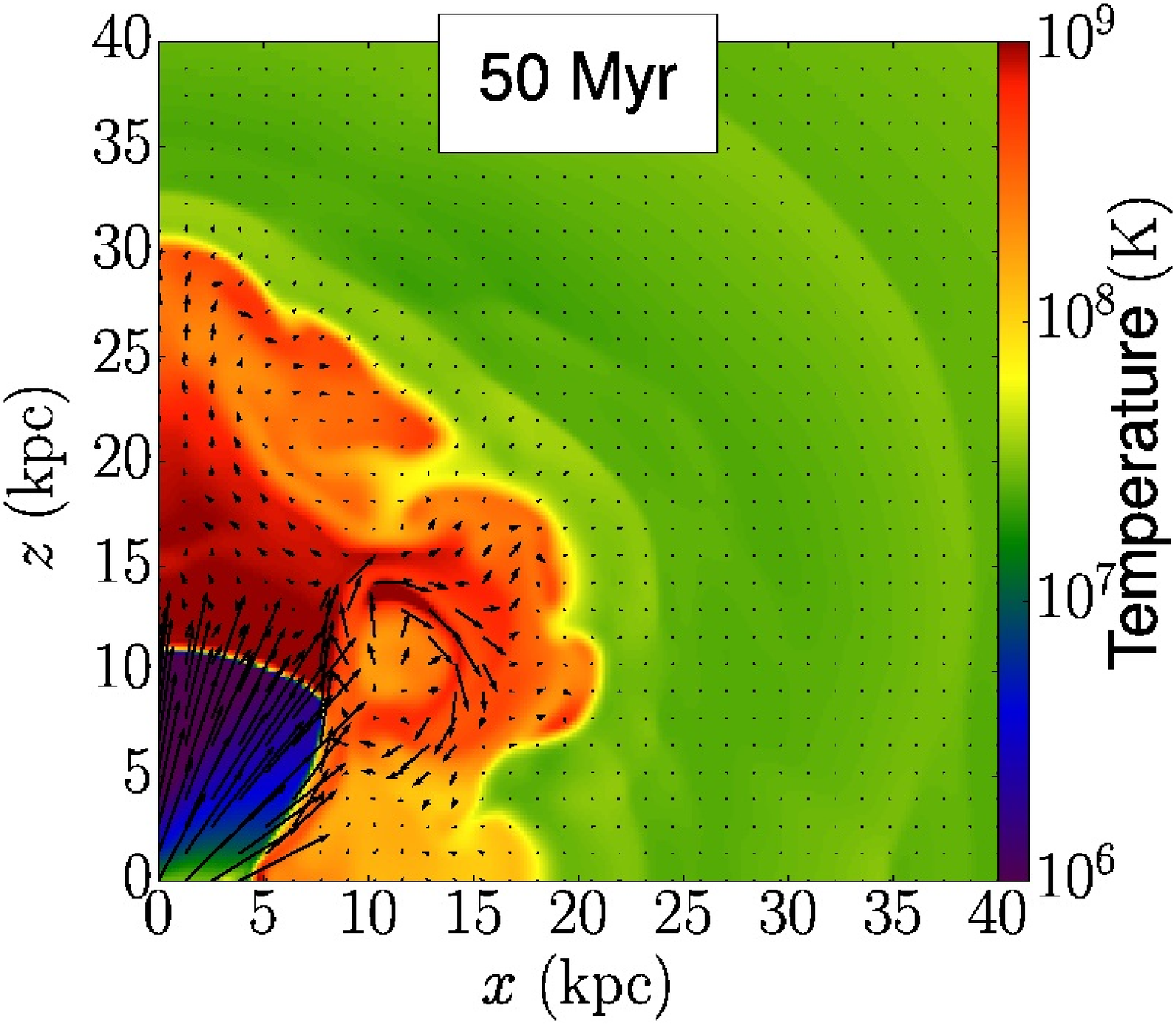}}
\subfigure{\includegraphics[width=0.32\textwidth]{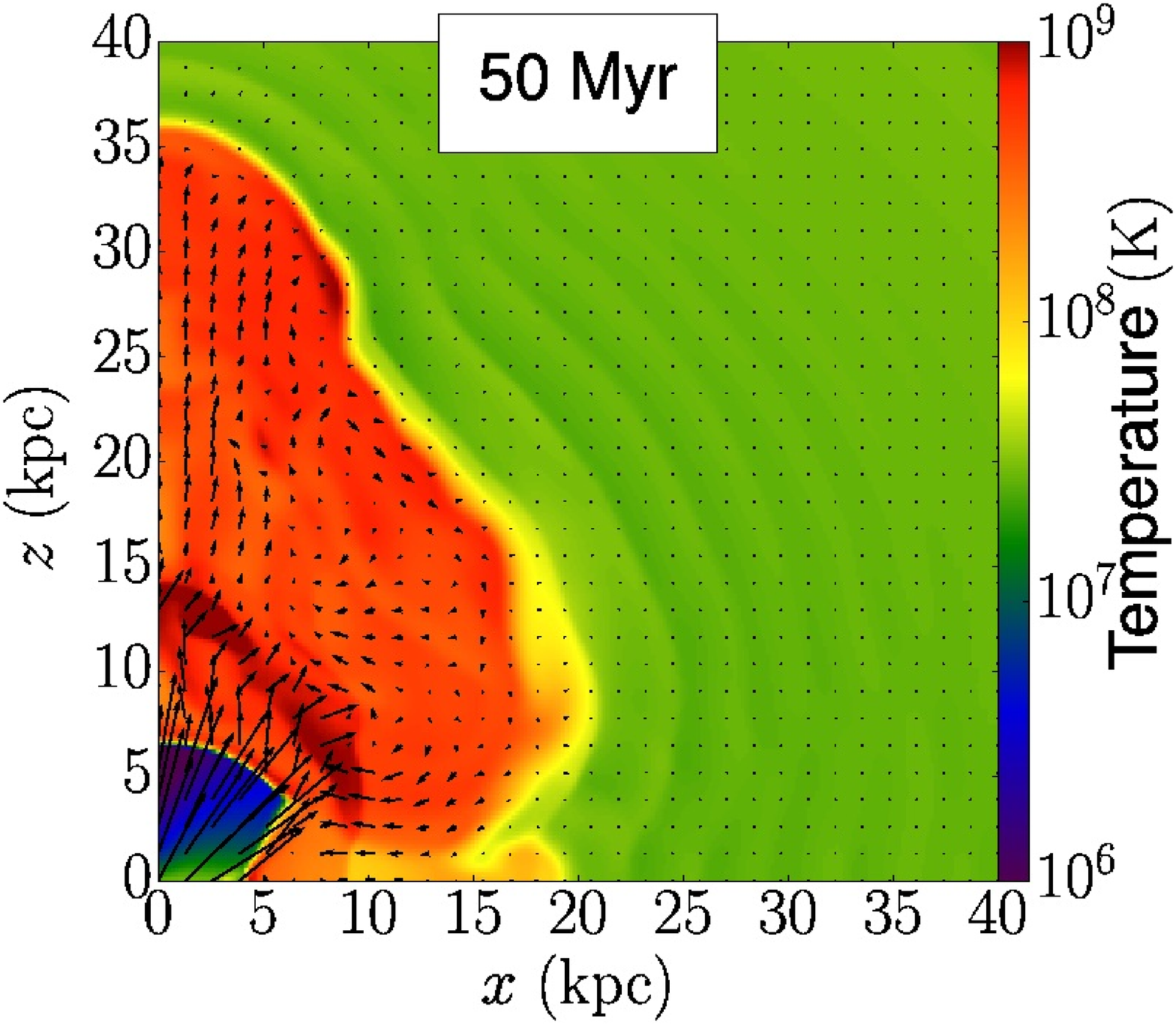}}
\subfigure{\includegraphics[width=0.32\textwidth]{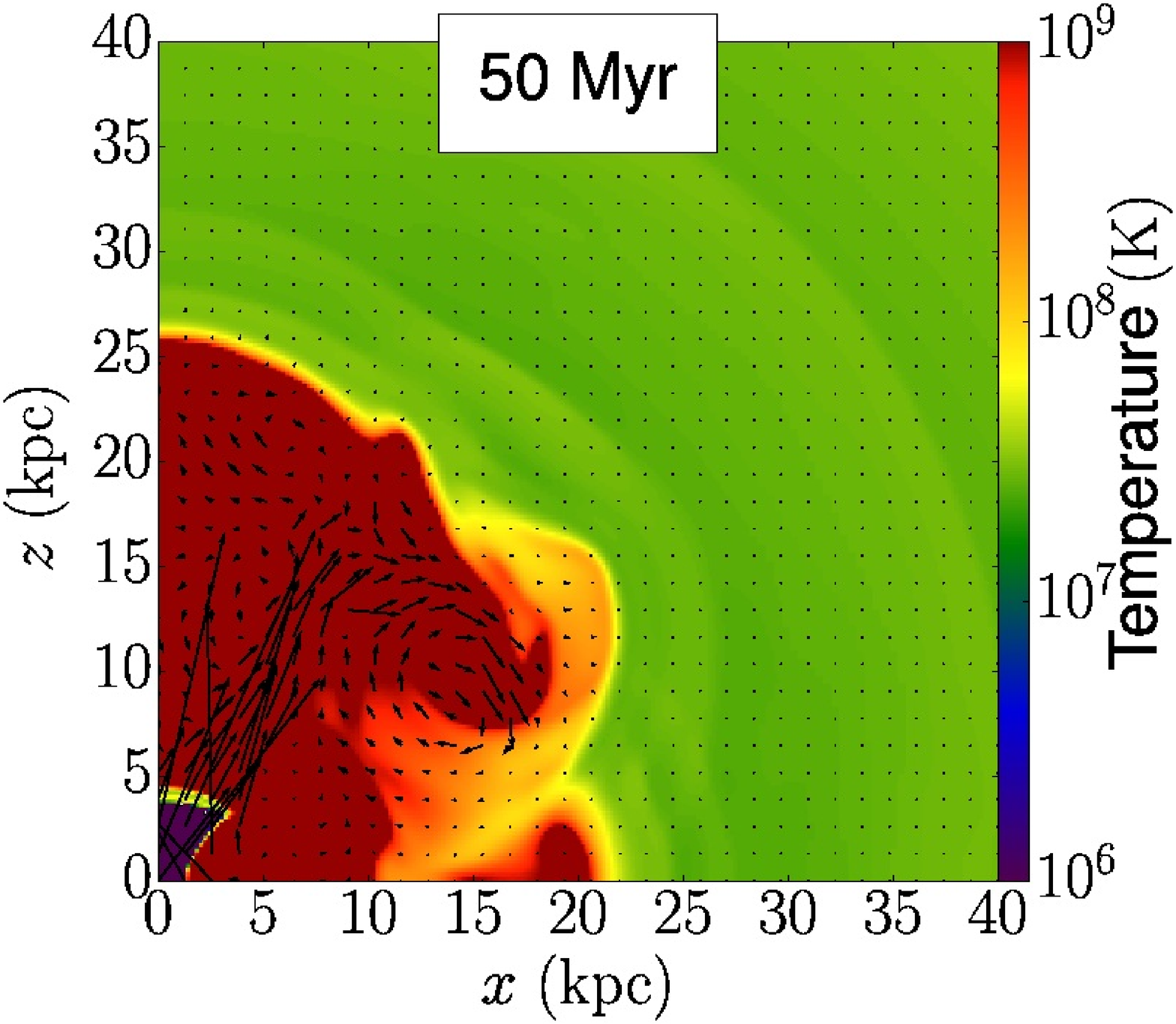}}
\caption{
Mass density and temperature maps of the nominal run (left column), Run B (middle column) and Run C (right column) whose energy histories are presented in Fig.~\ref{figure: energy history}.
The $y = 0$ slices of the simulations are presented at time $t = 50 \Myr$.
The color coding of the density is linear and presented in the range $(0.5, 1.5) \times 10^{-25} \g / \cm^3$, and the temperature coding is in logarithmic scale.
Arrows represent the velocity,
where a length of $1 \kpc$ on the map corresponds to $1700 \km \s^{-1}$.
}
\label{figure: additional simulations}
\end{figure}

We now turn to the additional two simulations presented in Fig.~\ref{figure: energy history}.
In one simulation, Run B, shown in the panels in the middle column of Fig.~\ref{figure: energy history}, the jet is periodically turned on for $2\Myr$ and off for
$2\Myr$, instead of $10 \Myr / 10 \Myr$ in the nominal run.
Density, temperature, and velocity maps in the $y=0$ plane at time $t=50\Myr$ of Run B are shown in the middle column of Fig.~\ref{figure: additional simulations}.
In Run C, shown in the panels in the right-hand column of Fig.~\ref{figure: energy history}, the jet periodicity is unchanged, but the mass deposition rate is decreased by a factor of $10$
and the jet velocity is increased by a factor of $\sqrt{10}$ with respect to the nominal simulation.
The rest of the parameters are as in the nominal simulation.
The flow maps for Run C in the $y = 0$ plane at t = 50 Myr are shown in the right-hand side column of Fig.~\ref{figure: additional simulations}.
In all simulations the mean jet power is the same, Eq.~(\ref{eq: jet power}).

The detailed flow and vortex structure in the three simulations is dependent upon the parameters in a complicated way.
However, we can draw conclusions which do not depend on the exact value of the parameters.
In Run B (middle column of Fig.~\ref{figure: energy history}), the time period of jet episode activity is $5$ times shorter.
The immediate effect seen is the smaller and more frequent `bumps' as a result of the shock waves.
In addition, the shape of the hot bubble is elongated in the direction of the axis of the jet, while maintaining a similar volume
which depends only on the jet energy output.
Run C, with the higher jet velocity and lower mass deposition rate, has a similar general structure to the nominal run,
but the details are different.
In Run C there is less momentum per unit mass and per unit energy, compared with the nominal run.
The bubble front along the $z$-axis after $50\Myr$ extends only to $27\kpc$, compared with $32\kpc$ in the nominal run.
The vortex on the right side of the bubble in Fig.~\ref{figure: additional simulations} is larger in Run C than in the nominal run.
The waist in Run C is wider than in the nominal run.
The nominal run better matches the morphology of observed bubbles.
This is one of the reasons we use slow jets \citep{Sternberg2007}.
Another reason is recent observations of such outflows from AGN (e.g., \citealt{Chamberlainetal2015}).

The energy histories of the traced regions in these simulations, Fig.~\ref{figure: energy history}, is similar to the nominal run
with the following exceptions.
In Run B (middle column of Fig.~\ref{figure: energy history}), in the tracer shown in the bottom panel,
the traced gas gets close to the hot bubble at $t=100-150\Myr$ but does not mix with the hot gas.
The more frequent episodes create less vortices and in fact push out the traced gas almost continuously.
In contrast, in the nominal run, where the jet is periodically turned on and off for $10\Myr$,
the longer quiescence time allows for the cooled ICM gas to be pulled in to the center and thus to heat up by mixing with the hot bubble gas. 
In Run B this happens to a lesser extent.
Thus, the traced ICM gas remains in the dense shell surrounding the hot bubble, which in turn radiates its energy more efficiently.
This radiative cooling is the cause of the decline of internal energy starting at time $t \approx 180\Myr$.
As seen in this example, more shock waves (keeping the jet power fixed) may, in some cases, actually heat some of the ICM gas less efficiently,
because it pushes the gas out more continuously, which inhibits mixing.

Similar cooling of the same traced region happens in Run C, as seen in the bottom-right panel of Fig.~\ref{figure: energy history}.
Here, the gas begins to mix at time $t \approx 110\Myr$, as in the nominal simulation.
However, the slightly different vortex structure does not pull in the entirety of the traced gas and most of it remains out of the hot bubble,
in the surrounding dense shell. Thus, as in Run B, it loses internal energy quickly by radiative cooling.
This may be a mechanism for feeding the AGN by cold gas.
ICM gas gets pulled in to the center during quiescence periods because of the lower pressure in the hot bubble,
cools radiatively when its density increases, and is then pulled in further.
At this stage it either heats up or cools further and falls in to feed the AGN.

The main results from the comparison of the different cases is as follows.
Because of the different flow structure in simulations with varying parameters,
some of the gas heats up more and some less, depending on the location of the gas,
but mixing is the main heating process among those probed in the simulations.

We thus conclude by arguing that mixing, due to vortices,
is more important as a heating process of the ICM gas by jets than turbulent heating and shock heating.
Shock waves induce an expansion and contraction and increase kinetic
energy by creating motion, but their contribution to the heating
is largely temporary and almost periodic in our setting. Turbulent
motion is also not the main heating mechanism, although it might
carry up to $\approx 0.2$ of the energy transferred from the jets
to the ICM.

\section{SUMMARY}
\label{s-summary}

Motivated by the recent claim of \cite{Zhuravlevaetal2014} that
turbulent-heating can counter radiative cooling in the cooling
flow clusters Virgo and Perseus, we compare the energy channeled
from jets to ICM turbulence with the direct thermal heating by
mixing hot bubble gas with the ICM. We have done so by conducting
3D hydrodynamical simulations of wide jets that inflate bubbles.
We used the {\sc pluto} \citep{Mignone2007} hydrodynamical code in
Cartesian coordinates.

The inflation of bubbles leads to the development of vortices in
the entire volume around the expanding jets and bubbles, as can be
see in  Figs.~\ref{figure: flow structure}-\ref{figure: tracers tr3} and \ref{figure: additional simulations}. These vortices excite turbulence in the ICM. These vortices
also efficiently mix ICM gas with the hot bubble gas. This mixing
can best be seen in Fig.~\ref{figure: tracers tr4} where the
tracer (colored red) of ICM gas located initially within a torus is
seen to be mixed with the bubble gas.

We use slow massive wide (SMW) outflows.
Such flows are supported by observational findings (e.g., \citealt{Chamberlainetal2015} and references therein).
In many cases outflows from AGN include relativistic components.
By comparing the results of Run C with the nominal run, e.g., in Fig. \ref{figure: additional simulations},
we note that when the velocity is three times as high, the bubble stays wide and a large vortex exists on its side.
Higher velocities do not change the mixing part by much when wide (fat) bubbles are inflated.
Therefore, we do not expect that relativist outflow will change our present conclusions, as long as wide bubbles are inflated.
For that, either the jets are wide, or there is a relative transverse velocity between the jets and the ICM, e.g., motion of the AGN relative to the ICM and/or jets' precession \citep{Sternberg2008a}.
Well-collimated relativistic jets without transverse velocity relative to the ICM will not heat the ICM efficiently away from their propagation cone.
They will penetrate to large distances.
They can form radio lobes at large distances, but will not heat the ICM close to the center.

We then compared the increase in thermal energy of the ICM medium
we trace, with the increase of its kinetic energy. We found that
the energy channeled to directly heating the gas is
more than 3 times larger than that channeled to kinetic energy of
the ICM, and typically larger than 4 times the kinetic energy.
This is our main result
and it is presented in Fig.~\ref{figure: energy history}. As only
part of the kinetic energy will turn to turbulence (some is just
large-scale motion), we conclude that
turbulent-heating is $\la 30\%$ as efficient as mixing-heating.
Heating by shocks is very small, as was shown
in 2D simulations \citep{GilkisSoker2012, HillelSoker2014}, and
was reinforced here in 3D simulations.
The significance of the new 3D simulations is that the vortices and turbulent motion is not artificially confined to be cylindrically symmetric.
Although the setup is cylindrically symmetric, small perturbations and numerical errors are magnified by hydrodynamic instabilities, and the flow is fully three-dimensional.
This is the main difference observed in the 2D and 3D simulations.
The conclusions regarding the heating mechanisms, nonetheless, are similar.

It is important to note that since by our finding turbulence
carries
$\la 25 \%$
of the energy deposited by the jets in
the inner region,
turbulence may still be non-negligible in the
ICM, as was found recently by \cite{Zhuravlevaetal2014},
\cite{Zhuravlevaetal2015}, \cite{Arevalo2015}, and marginally by \cite{AndersonSunyaev2015}.
However, since only a portion of the kinetic energy develops to turbulence,
and since the total injected jet energy can be channeled into other forms of energy,
in some cases turbulence might be insignificant.

\cite{Zhuravlevaetal2014} find observationally that turbulence is present and
significant in the ICM, and that the rate of turbulent heating matches the
rate of radiative cooling. In contrast, in our simulations we have found that heating by mixing is more
significant.
Our results further show that a non-negligible amount of energy ends up as
kinetic energy of the ICM. However, substantial heating occurs only when
mixing takes place. In addition, we find large density fluctuations induced in
the ICM by the mixing process and bubble motion; this is best seen in the upper panels of Fig. \ref{figure: additional simulations}.
It seems that a large
fraction of the kinetic energy is not turbulent motion that dissipates. The
kinetic energy and density fluctuations are associated also with the bubble
activity and global gas motion; only part of of the kinetic energy belongs
to the universal turbulent cascade that \cite{Zhuravlevaetal2014} refer to.
\cite{Zhuravlevaetal2014} write: ``It is difficult to prove unambiguously
that we are dealing with a universal turbulent cascade, as other structures
(e.g., edges of the bubbles, entrainment of hot bubble matter, sound waves,
mergers and gas sloshing) might also contribute to the observed
density-fluctuation spectra.''
We suggest that the discrepency between the claim of
\cite{Zhuravlevaetal2014} for efficient turbulent heating and our finding
results from a large fraction of the density perturbations and turbulent energy that is indeed in the
other structures listed by them. This definitely deserves further
investigation.

We thank an anonymous referee for very valuable suggestions and comments.

\end{document}